# Optimal individualized treatment regimes for survival data with competing risks

Christina W. Zhou
Department of Biostatistics, University of North Carolina at Chapel Hill
Nikki L. B. Freeman,
Department of Biostatistics and Bioinformatics, Duke University
Katharine L. McGinigle,
Division of Vascular Surgery, UNC-CH
and
Michael R. Kosorok
Department of Biostatistics, UNC-CH

November 5, 2024

**Abstract**

Precision medicine leverages patient heterogeneity to estimate individualized treatment regimens—formalized, data-driven approaches designed to match patients with optimal treatments. In the presence of competing events, where multiple causes of failure can occur and one cause precludes others, it is crucial to assess the risk of the specific outcome of interest, such as one type of failure over another. This helps clinicians tailor interventions based on the factors driving that particular cause, leading to more precise treatment strategies. Currently, no precision medicine methods simultaneously account for both survival and competing risk endpoints. To address this gap, we develop a nonparametric individualized treatment regime estimator. Our two-phase method accounts for both overall survival from all events as well as the cumulative incidence of a main event of interest. Additionally, we introduce a multi-utility value function that incorporates both outcomes. We develop random survival and random cumulative incidence forests to construct individual survival and cumulative incidence curves. Simulation studies demonstrated that our proposed method performs well, which we applied to a cohort of peripheral artery disease patients at high risk for limb loss and mortality.

# 1 Introduction

Precision medicine (PM) aims to improve healthcare decision-making by tailoring medical treatments to individual characteristics (Kosorok and Moodie, 2015; Jiang et al., 2017; Moodie and Krakow, 2020; Kosorok et al., 2021). One way to formalize this goal is through individualized treatment regimes (ITRs), maps from patient characteristics to treatment recommendations (Kosorok and Laber, 2019). Ideally, we would like to identify optimal ITRs, the treatment regimes that, if followed, would yield the best outcome on average in the targeted population (Kosorok and Laber, 2019; Kosorok and Moodie, 2015). One way to evaluate this is through the value function, the expected clinical outcome based on a specified treatment regime.

In healthcare settings, identifying the best outcome can be challenging, especially when managing complex health conditions. For example, chronic limb threatening ischemia (CLTI), the most severe form of peripheral artery disease (PAD), is a disease characterized by abnormal narrowing and occlusion of the arteries in the lower extremities. CLTI is a marker for severe, poorly controlled cardiovascular disease and diabetes. Even with medical and surgical treatment, patients with CLTI are at high risk for devastating outcomes, including limb loss, cardiac events (e.g., heart attack and stroke), and death (Criqui et al., 2021). From a statistical perspective, CLTI and CLTI outcomes demonstrate an example of competing risks (CR). Competing risks arise when individuals are at risk for multiple types of events, and the occurrence of one event precludes or alters the probability of other events occurring. For example, if a patient with CLTI dies, they are no longer at risk for limb loss. Standard survival analyses ignore CR by including patients experiencing the competing risk as part of the risk set, despite that they can no longer experience the main event of interest. Thus, methods estimating the cumulative incidence function (CIF), or subdistribution, is often of interest when analyzing CR data (Kalbfleisch and Prentice, 1980; Fine and Gray, 1999; Austin et al., 2016). By accounting for competing risks, estimation methods avoid inflation or bias in event probabilities or survival estimates.

Patients with CLTI are highly heterogeneous, and evidence suggests that precision medicine strategies may improve treatment recommendations and, ultimately, patient outcomes (McGinigle et al., 2021; Chung et al., 2022). Current ITR estimation methods for handling the competing risk setting, like that posed by CLTI, focus on either optimizing the survival curve or minimizing risks of individual events, i.e., minimizing the CIF. However, none of the existing methods *simultaneously* optimize the survival function and the CIF.

We address this gap by proposing a nonparametric two-phase optimal ITR estimator that leverages multi-utility optimization. To do this, we construct a novel multi-utility value (objective) function that maximizes overall survival (OS) from all causes while minimizing the CIF of a priority cause (PC), the main event of interest over any other cause of failure. The first phase of our estimation strategy constructs individual OS curves by adapting the generalized random survival forests from Cho et al. (2023) to the single stage treatment setting. We develop random cumulative incidence forests (RCIFs) to construct individualized CIF curves for the PC in the second phase. To select an optimal ITR, we use a flexible, user-specified criterion (i.e., a tolerance level) to balance treatments so that they appropriately weigh survival and risks. Unlike most existing methods restricted to binary treatments, our method considers a flexible finite number of treatment arms. Because our forests are nonparametric, we can capture complex decision rules. The software package implementing our proposed method is made freely available at https://github.com/cwzhou/itrSurv.

The rest of this paper is organized as follows: Section 2 reviews existing literature for precision medicine and the competing risk setting. Section 3 details our analytic framework and introduces our proposed method. Section 4 presents the theoretical properties of our ITR estimator and forests. Section 5 describes the finite sample performance of our proposed estimator through simulation studies. We demonstrate our proposed method's application to the CLTI setting in Section 6 and follow with a discussion in Section 7.

# 2 Background

In this section, we briefly review precision analytic survival and CR methods. We conclude this section with foundational survival and CR ideas that we will use through the remainder of the manuscript.

## 2.1 PM methods for survival and CR outcomes

Our work uses tree-based estimators to construct ITRs that simultaneously account for survival outcomes and CRs. Although a variety of innovative methods have been developed for estimating ITRs with survival outcomes (Zhao et al., 2015; Hager et al., 2018; Wu et al., 2020; Zhong et al., 2021; Schrod et al., 2022; Choi et al., 2023; Bakoyannis, 2023), only a few have employed tree-based methods or accounted for multiple outcomes. Forest-based methods include Cui et al. (2017); Zhu et al. (2018); Doubleday et al. (2022); Devaux et al. (2023). Recently, Cho et al. (2023) introduced the generalized random survival forest, a tree-based approach to predict individualized survival curves. The predicted survival curves are then used to estimate optimal multi-stage ITRs that maximize the survival curve. Our proposed method extends the work of Cho et al. (2023) by incorporating multiple outcomes, i.e., OS and PC cumulative incidence.

Current precision medicine methods for CR is limited. Yavuz et al. (2018) proposed a nonparametric CIF estimator that was then used to construct optimal ITRs that minimize the CIF using SMART data. Zhou et al. (2021) focused on minimizing side effects in a CR setting by estimating the optimal ITR at a specific time point, with side effects constrained below a set threshold. In He et al. (2021), the authors minimize the CIF to identify the optimal ITR using a doubly robust method to guard against model misspecification, which ensures consistency if either model component is correctly specified.

In terms of multiple outcomes, Liu et al. (2023) introduced a multi-stage cumulative benefit-risk framework for sequential multiple assignment randomized trials (SMARTs). By incorporating a trade-off framework, Liu et al. (2023) maximized the expected benefit while simultaneously keeping the expected cumulative risk below a pre-defined threshold. This framework is not designed to address the complexities of the CR setting, where multiple failure types may occur.

Standard practices for multiple endpoints include developing composite outcomes (Neuhauser, 2006; Jetsupphasuk et al., 2023) or using multivariate models (Teixeira-Pinto et al., 2009; Wu et al., 2024). However, composite outcomes can be difficult to interpret, especially if one component drives the overall result. An alternative approach is maximizing an utility function, a fixed summary measure of outcomes (Hayes et al., 2022). Utility functions for multiple outcomes, which we call multi-utility, have been used for developing

PM algorithms (Butler et al., 2018; Jiang et al., 2021; Zitovsky et al., 2023). However, existing methods are not designed for survival data, where censoring may occur. To our knowledge, there are no existing precision medicine methods for survival outcomes with CRs that simultaneously account for both outcomes, and there are no existing precision medicine methods for survival data that utilize a multi-utility value function.

## 2.2 Foundational Survival and CR ideas

We briefly review key concepts for survival analysis and CRs. Let $T$ represent failure time and $Z$ represent the associated covariates. The survival function is the probability of not experiencing an event over time, while the hazard rate is the instantaneous risk of an event occurring at a specific time, given that the event has not occurred up to that point. The OS function is $S(t; Z) = Pr(T > t|Z) = \exp\left[-\int_0^t \lambda(u; Z)du\right]$, where $\lambda(t; Z)$ is the hazard function at time $t$.

The CIF, or subdistribution, is the probability that a specific event has occurred by a certain time, while considering the overall risk set of individuals, including the potential occurrence of other competing events. The CIF differs from the traditional survival function, which focuses on time to first event regardless of type. Let $f_j(t)$ be the (sub)density function for the time to a cause $j$ failure, then the cause-$j$ specific hazard is defined as $\lambda_j(t; Z) = d\Lambda_j(t; Z) = \frac{f_j(t;Z)}{S(t;Z)}$ (Kalbfleisch and Prentice, 1980). A jump point exists when an individual experiences an event of interest, and the probability of this event occurring at the jump point is $d\Lambda_j(t; Z) = Pr(T = t, \epsilon = j|T \geq t; Z)$. If $t$ is a continuity point of $\Lambda_j(t; Z)$, then $d\Lambda_j(t; Z) = \lambda_j(t; Z)dt$, indicating a smooth and consistent risk over a very small time interval around $t$. The CIF for cause $j$ corresponding to the external covariate as $F_j(t; Z) = Pr(T \leq t, J = j|Z) = \int_0^t S(u - |Z)d\Lambda_j(u; Z)$, where $S(t-)$ is the left-hand limit of the OS function at time $t$ and $\Lambda_j(t)$ is the (left continuous) cumulative cause $j$ specific hazard function (Kalbfleisch and Prentice, 1980).

# 3 Proposed Methods

## 3.1 Framework and Notation

Let $\epsilon_i$ be the cause indicator such that $\epsilon_i = j$ means individual $i$ experienced failure from cause $j$, where $j = 1, \ldots, J$. Our focus is on a priority cause of interest. In cases with more than two causes, we collapse the causes into two such that $Pr(\epsilon = 1) + Pr(\epsilon = 2) = 1$: the PC (cause 1) and all others (cause 2). Let $T_{1i}$ and $T_{2i}$ denote the failure times for causes 1 and 2, respectively, for patient $i$, and $T_i = \min(T_{1i}, T_{2i})$ be the overall survival time, representing time to the first failure. Right-censoring due to study end ($\tau$) or loss to follow-up is denoted $C_i$, with censoring time distribution $G_i(t) = \Pr(C_i \geq t|Z_i)$. The observed time is $X_i = \min(T_i, C_i)$. The OS event indicator is $\delta_i = I(T_i \leq C_i)$. We define $\delta_{1i} = I(T_{1i} \leq C_i \wedge T_{2i})$ and $\delta_{2i} = I(T_{2i} \leq C_i \wedge T_{1i})$.

We assume a finite treatment space $\mathcal{A}$. Let $Z_i = (Z_{i1}, \ldots, Z_{iP})^T$ represent the $P$ covariates. For each $z \in Z$, let $\psi(z) \subseteq A$ to be the set of allowable treatments for a patient with $Z = z$. We observe data composed of $n$ i.i.d replicates $\mathcal{D}_n = \{Z_i, X_i, \delta_i, \delta_{1i}\}_{i=1}^n$. When unambiguous, we drop the index $i$ for simplicity.

Let $d$ represent an ITR, which maps patient covariate information ($z_i \in Z$) to the treatment assignment ($a_i \in \mathcal{A}$), $d(z_i) : z_i \to a_i$ for each patient $i$, which satisfies $d(z) \in \psi(z)$ for all $z \in Z$. Let $\phi(\cdot)$ be an optimization criteria for OS and CI. In this manuscript, we focus on the truncated (at $\tau$) area under the curves, $\phi(S) = \int_0^\tau S(t)\,dt$ and $\phi(F_j) = \int_0^\tau F_j(t)\,dt$, the restricted mean survival time at time $\tau$ and integrated CIF truncated at $\tau$ respectively. The latter quantity measures the accumulation of the probability of having the primary event of interest during the follow-up period.

For any treatment $a \in \mathcal{A}$ and covariates vector $Z$, let $S^{(a)}(\cdot \mid Z)$ and $F_j^{(a)}(\cdot \mid Z)$ be the potential survival and cumulative incidence curves for patients with covariates $Z$ and treatment $a$, respectively. Let $\phi_1^{(a)}(Z) = \phi(S^{(a)}(\cdot \mid Z))$ and $\phi_2^{(a)}(Z) = \phi(F_j^{(a)}(\cdot \mid Z))$ denote the potential outcomes of interest. Under regime $d$, define the potential outcomes as $\phi_1^{(d)}(Z) = \sum_{a\in\mathcal{A}} \phi_1^{(a)}(Z)\mathbf{1}\{d(Z) = a\}$ and $\phi_2^{(d)}(Z) = \sum_{a\in\mathcal{A}} \phi_2^{(a)}(Z)\mathbf{1}\{d(Z) = a\}$, where $\mathbf{1}\{B\}$ is the indicator of $B$. We assume that a prolonged time to failure is optimal, so we optimize a criterion ($\phi_1$) to maximize OS. Additionally, we want to minimize the CIF of the PC via $\phi_2$ due to its inherent importance over all other causes. We aim to estimate the optimal ITR $d_{\text{opt}}$ that optimizes a joint criterion $\boldsymbol{\phi}^{(d)}(Z) = \begin{pmatrix}\phi_1^{(d)}(Z)\\ \phi_2^{(d)}(Z)\end{pmatrix} = \sum_{a\in\mathcal{A}} \begin{pmatrix}\phi_1^{(a)}(Z)\\ \phi_2^{(a)}(Z)\end{pmatrix} \cdot \mathbf{1}\{d(Z) = a\}$. For any treatment regime $d(Z)$, we define the joint value function of the ITR estimator as

$$\mathbb{V}(d) = \sum_{a\in\mathcal{A}} \mathbb{E}\left[\boldsymbol{\phi}^{(a)}(Z) \cdot \mathbf{1}\{d(Z) = a\}\right].$$

Define $a_1^* = \arg\max_{a\in\mathcal{A}} \phi_1^{(a)}(Z)$ with $V_1^*(Z) = \phi_1^{(a_1^*)}(Z)$. If OS across treatments are sufficiently close (i.e., within a tolerance level $\alpha_\phi$), then for a treatment $a$,

$$1 - \frac{\phi_1^{(a)}(Z)}{V_1^*(Z)} \le \alpha_\phi. \tag{1}$$

This condition assesses if the treatment $a$ performs nearly as well as the best available treatment within the acceptable margin. Equation (1) implies $\phi_1^{(a)}(Z) \ge (1 - \alpha_\phi)V_1^*(Z)$. Thus, the optimal ITR is defined as $d_{\text{opt}}(Z) = \begin{cases}\arg\min_{a\in\mathcal{A}} \phi_2^{(a)}(Z) & \text{if } \phi_1^{(a)}(Z) \ge (1-\alpha_\phi)V_1^*(Z),\\ \arg\max_{a\in\mathcal{A}} \phi_1^{(a)}(Z) & \text{otherwise.}\end{cases}$

## 3.2 Proposed Estimators

To identify the potential outcomes, we assume Assumptions 1-3 (Kidwell, 2016; Kosorok and Laber, 2019). Then, using identified data, the value function can be written as $\mathbb{V}(d) = \sum_{a\in\mathcal{A}} \mathbb{E}\left[\boldsymbol{\phi}(a, Z) \cdot \mathbf{1}\{d(Z) = a\}\right]$, and we want to estimate the conditional probabilities $\phi_1(a, Z) = \int_0^\tau S(t|Z, A = a)\,dt$ and $\phi_2(a, Z) = \int_0^\tau F_j(t|Z, A = a)\,dt$.

**Assumption 1** (Stable Unit Treatment Value Assumption (SUTVA))**.** *Each individual's counterfactual outcomes, such as cause-specific failure time or overall failure time, are not affected by the treatment assignments or histories of other individuals in the study. Additionally, the process of censoring is assumed to be independent of other individuals' treatments and histories.*

**Assumption 2** (Strong ignorability). *The treatment assignment is conditionally independent of future potential outcomes (such as $\phi_1^{(a)}$ or $\phi_2^{(a)}$), given the individual's covariates $Z$. This can be expressed as $\phi^{(a)} \perp A \mid Z$, for all $a \in A$.*

**Assumption 3** (Positivity). *Given covariate information, the probability of having each treatment $a \in \mathcal{A}$ is greater than a constant $L > 0$, or $\Pr(A = a|Z = z) > L$ for all $a \in \mathcal{A}$, $z$ such that $\Pr(Z = z) > 0$.*

### 3.2.1 Random Forests

To estimate individualized curves for OS and PC cumulative incidence, we develop a random survival forest (RSF) and random cumulative incidence forest (RCIF), respectively. We adapt the generalized random survival forest in Cho et al. (2023) to the single stage treatment setting. We provide details on the construction of the proposed RSF in the Supplementary Material.

Following the construction of the RSF, we develop RCIFs to obtain individual CIF curves for optimization (see Algorithm 1). The Aalen-Johanson (AJ) estimator $\widehat{F}_j$ is used to estimate the cumulative incidence function, as well as for node splitting. Letting $k$ represent the node, the AJ estimator for the CIF for cause $j$ failure (i.e., $\epsilon = j$) in group $k$ is given by:

$$\widehat{F}_{jk}(t) = \int_0^t \widehat{S}_k(u-)Y_k^{-1}(u)dN_{jk}(u) \tag{2}$$

where $\widehat{S}_k(t-)$ is the left-hand limit of the Kaplan-Meier estimate $\widehat{S}_k(t)$ for OS (see Equation (S1) in Supplementary Material), $N_{jk}(t)$ is the number of failures of cause $j$ by $t$, $Y_k(t)$ is the number of subjects still at risk just prior to $t$ in group $k$ (Gray, 1988). In building the trees for the RCIF, $k$ denotes the two groups that are being compared; however, for prediction, $k = 1, \dots, K$ indexes the terminal node, where $K$ is the number of terminal nodes. Whenever $F_{jk}(t)$ refers to the CIF within terminal nodes, we omit the index $k$ such that $F_{jk}(t) \equiv F_j$, since there is no risk of ambiguity.

In order to partition a node into two potential daughter nodes $(\mathcal{L}_1, \mathcal{L}_2)$, we use Gray's two-sample test (Gray, 1988) to compare the priority cause CIF between groups:

$$\text{Gray}(\mathcal{L}_1, \mathcal{L}_2) = \int_0^\tau K(t)\{[1 - \widehat{F}_{11}(t-)]^{-1}d\widehat{F}_{11}(t) - [1 - \widehat{F}_{12}(t-)]^{-1}d\widehat{F}_{12}(t)\}, \tag{3}$$

where $\widehat{F}_{1k}$ is an estimate of $F_{1k}$ based on the AJ estimator for CIF of failure cause 1 in group $k$ and $K(t)$ is some weight function. In this paper, we define $K(t) = 1$. The splitting rule chooses $(\mathcal{L}_1, \mathcal{L}_2) = \arg\max_{\mathcal{L}_1,\mathcal{L}_2} \text{Gray}(\mathcal{L}_1, \mathcal{L}_2)$ as the best partition. The recursive partitioning continues with a minimum of $n_{\text{minevent}}$ events (from any cause) in each node while each node's size exceeds the threshold $n_{\min}$. The PC cumulative incidence estimate of the terminal node $\mathcal{L}$ of the $w$-th cumulative incidence tree is given by the AJ estimator $\widehat{F}_{j,\{w\}}(t \mid \mathcal{L})$, where $w = 1, 2, \dots, n_{\text{tree}}$. Letting $h = (a, Z)$, the RCIF estimate for priority cause $j$ is given by

$$\widehat{F}_j(t \mid h) = \frac{1}{n_{\text{tree}}} \sum_{w=1}^{n_{\text{tree}}} \sum_{l=1}^{\#\{\mathcal{L}_{\{w\},l}:l\}} 1(h \in \mathcal{L}_{\{w\},l})\widehat{F}_j(t \mid \mathcal{L}_{\{w\},l}),$$

where $\#\{\mathcal{L}_{\{w\},l} : l\}$ is the number of terminal nodes of the $w$-th tree.

---

**Algorithm 1** The random cumulative incidence forest estimator

**Input:** $\{(X_i, \delta_i, \delta_{ji}, Z_i, A_i)\}_{i=1}^n$

**Output:** $\widehat{F}_j(\cdot|Z, A = a)$

Result: random cause-j cumulative incidence forest $\widehat{F}_j(\cdot|Z, A)$

Parameters: number of trees $n_{\text{tree}}$, minimum terminal node size $n_{\min}$, minimum number of events $n_{\text{minevent}}$

For treatment arm $a \in \mathcal{A}$ **do**

For tree $w = 1, 2, \ldots, n_{\text{tree}}$ **do**

Partition the feature space with the input data $\{(X_i, \delta_i, \delta_{ji}, Z_i)\}_{i:a_i=a}$ via Gray's test (3) until node size $< 2n_{\min}$ and there are at least $n_{\text{minevent}}$ events from any cause in every node

Obtain the terminal node cumulative incidence probability $\widehat{F}^a_{j,\{w\}}(t|Z)$ via Aalen-Johanson estimator (2)

Obtain the random cumulative incidence forest estimator $\widehat{F}_j(\cdot|Z, A = a) = \frac{1}{n_{\text{tree}}} \sum_{w=1}^{n_{\text{tree}}} \widehat{F}^a_{j,\{w\}}(t|Z)$

---

#### 3.2.2 Estimation Procedure

We propose a two-phase estimation procedure that prioritizes maximizing the criterion of OS in Phase 1 but also minimizes the criterion of the CIF of a priority cause in Phase 2 over all treatments for each individual. Whether an individual's treatment is obtained from Phase 1 or Phase 2 depends on a pre-specified tolerance level $0 < \alpha_\phi < 1$.

Let $\widehat{S}(\cdot; a, Z)$ and $\widehat{F}_j(\cdot; a, Z)$ denote the estimated proposed RSF and RCIF forest estimators, respectively. Letting the subscript denote the method phase, we define the true and estimated functions of $\phi_1(a, Z)$ and $\phi_2(a, Z)$ as $\boldsymbol{\phi}^0(a, Z) = \begin{pmatrix} \phi_1^0(a, Z) \\ \phi_2^0(a, Z) \end{pmatrix}$ and $\widehat{\boldsymbol{\phi}}(a, Z) = \begin{pmatrix} \widehat{\phi}_1(a, Z) \\ \widehat{\phi}_2(a, Z) \end{pmatrix}$, respectively. Let the true optimal treatment for Phase 1 be $a_1^{0*} = \arg\max_{a \in \mathcal{A}} \phi_1^0(a, Z)$ with $V_1^{0*}(Z) = V_1^0(a_1^{0*}, Z)$, and the estimated value as $\widehat{a}_1^* = \arg\max_{a \in \mathcal{A}} \widehat{\phi}_1(a, Z)$ with $\widehat{V}_1^*(Z) = \widehat{V}_1(\widehat{a}_1^*, Z)$. We can then use the following two-phase procedure to estimate the optimal ITR. For each patient:

(i) estimate the conditional overall survival $S(t|Z, A)$ for each treatment

(ii) estimate the conditional priority j-cause CIF $F_j(t|Z, A)$ for each treatment

(iii) Phase 1: optimize (i) over feasible finite treatment space for all $a \in \psi(z) \subset \mathcal{A}$ to obtain optimal treatment for overall survival

(iv) Find $A^s$, the set of treatments chosen using $\alpha_\phi$ criteria to compare to the optimal overall survival treatment

   (a) if the cardinality of $A^s$ is 1 then we have ITR estimate $\widehat{d}(z) = A^s$

(b) if the cardinality of $A^s$ is greater than 1 then optimize (ii) (Phase 2) over treatment set $A^s$ to obtain ITR estimate $\widehat{d}(z)$.

---

**Algorithm 2** The proposed ITR estimator accounting for competing risks.

**Input:** $Z_i, A_i, \alpha_\phi$

**Output:** $\widehat{d}_{\text{opt}}$

For each individual $i$, given covariates $Z_i$ and small $0 < \alpha_\phi < 1$, consider feasible treatments $a_i \in \psi(z_i) \subset \mathcal{A}$ (we drop $i$ for simplicity):

1. Obtain $\widehat{S}(\cdot|Z, A)$ via random survival forest (Cho et al., 2023)
2. Obtain $\widehat{F}_j(\cdot|Z, A)$ via random cumulative incidence forest (Algorithm 1)
3. Let the estimated optimal treatments for Phase 1 (overall survival) be $\hat{a}_1^* = \arg\max_{a\in\psi(z)} \widehat{\phi}_1(a, Z)$ and $\widehat{V}_1^*(Z) = \widehat{V}_1(\hat{a}_1^*, Z)$
4. Obtain overall survival treatment set $A^s = \{a \in \psi(Z) : \widehat{\phi}_1(a, Z) \geq (1 - \alpha_\phi)\widehat{V}_1^*(Z)\}$
   - If $A^s$ is singleton, $A^s$ is the estimated optimal treatment: $\widehat{d}_{\text{opt}}(Z) = A^s$.
   - If $\#A^s > 1$, optimal treatment for Phase 2 is $\widehat{d}_{\text{opt}}(Z) = \arg\min_{a\in A^s} \widehat{\phi}_2(a, Z)$.

In other words, if $a_i \neq A_i^s$ for individual $i$, then individual $i$ stopped at either

Phase 1: $\widehat{\phi}_1(a, Z) < (1 - \alpha_\phi)\widehat{V}_1^*(Z)$ or

Phase 2: $\widehat{\phi}_1(a, Z) \geq (1 - \alpha_\phi)\widehat{V}_1^*(Z)$ but $\widehat{\phi}_2(a = \widehat{d}_{\text{opt}}, Z) < \widehat{\phi}_2(a = a_i, Z)$

and $A_i^s$ may have $\#A_i^s > 1$.

---

Algorithm 2 provides a high-level outline of our proposed method. Briefly, if the areas under the truncated survival curve between the treatments are within certain $\alpha_\phi$, then we want to reduce PC cumulative incidence. We consider the competing risk of a priority cause if the OS experience is not too different. However, if the OS experience is excessively different, then we proceed as if the competing risk is not relevant. From Equation (1), $1 - \frac{\widehat{\phi}_1(a,Z)}{\widehat{V}_1^*(Z)} \leq \alpha_\phi \implies \widehat{\phi}_1(a, Z) \geq (1 - \alpha_\phi)\widehat{V}_1^*(Z)$. Thus, the optimal ITR when $\#A^s > 1$ is defined as $\widehat{d}_{\text{opt}}(Z) = \arg\min_{a\in\mathcal{A}:\widehat{\phi}_1(a,Z)\geq(1-\alpha_\phi)\widehat{V}_1^*(Z)} \widehat{\phi}_2(a, Z)$.

# 4 Theoretical Properties

We show consistency of our proposed optimal ITR estimator $\widehat{d}_{\text{opt}}(Z)$, as well as the random survival forest in the single stage setting and our proposed RCIF. The proofs of all theorems are presented in the Supplementary Material. We use the definition of regular trees and random-split trees from the random forest literature (Meinshausen, 2006; Wager and Walther, 2016; Cui et al., 2022).

**Definition 1** (Random-split and $\alpha$-regular trees and forests)**.** *A tree is called a random-split tree if each feature is given a minimum probability ($\psi/d$) of being the split variable at each intermediate node, where $0 < \psi < 1$ and $d$ is the dimension of the feature space. A* *tree is $\alpha$-regular if every daughter node has at least an $\alpha$ fraction of the training sample in the corresponding parent node. A random forest is called a random-split ($\alpha$-regular) forest if each of its member trees is random-split ($\alpha$-regular).*

In addition to Assumptions 1-3 in Section 3.2, we also assume the following assumptions. Assumption 4 is important for asymptotic results, and is typical for survival analysis (Fleming and Harrington, 2011; Cui et al., 2017). This assumption ensures we only calculate each survival or cumulative incidence curve up to a time point with sufficiently large sample size.

**Assumption 4.** *There exists a fixed maximum follow-up time $0 < \tau < \infty$ and a constant $M \in (0, 1)$, such that $Pr(X_\tau \geq \tau | Z = z) \geq M$, for all $z \in \mathcal{Z}$.*

Below are assumptions used for consistency of both forests. Assumptions 5-7 are necessary and realistic for the tree-based estimator terminal node size and splitting rules. Assumption 5 requires the terminal node size to be not only sufficiently large absolutely but also sufficiently small relative to the sample size. Assumption 6 follows from Definition 1. Assumption 7 adds smoothness to the survival function and censoring cumulative distribution function. Assumption 8 follows from Cui et al. (2017) and Wager and Walther (2016), which allows for dependency between covariates. Thus, non-hyper-rectangular or categorical history spaces are allowed with this assumption. Assumption 9 is necessary for the random cumulative incidence forest because there is non-differentiability of the treatment policy.

**Assumption 5** (Terminal node size, polynomial)**.** *The minimum size $n_{min}$ of the terminal nodes grows at the rate $n_{min} \bowtie n^\beta$, $\frac{1}{2} < \beta < 1$, where $a \bowtie b$ implies that both $a = O(b)$ and $b = O(a)$.*

**Assumption 6** ($\alpha$-regular and random-split trees)**.** *Trees are $\alpha$-regular and random-split with a constant $0 < \psi < 1$.*

**Assumption 7** (Lipschitz continuous and bounded survival and censoring cumulative densities)**.** *There exist constants $L_S$ and $L_G$ such that $|S(t|h_1) - S(t|h_2)| \leq L_S \|h_1 - h_2\|$ and $|G(t|h_1) - G(t|h_2)| \leq L_G \|h_1 - h_2\|$ for all $h_1, h_2 \in \mathcal{H}$, $t \in [0, \tau]$, where $G$ is the censoring cumulative distribution function (CDF) and $S(\tau^- \mid h)G(\tau^- \mid h) > c_1$ for all $h$ and some constant $c_1 > 0$.*

**Assumption 8.** *The covariates $Z \in [0; 1]^d$ are distributed according to a density function $p(\cdot)$ satisfying $1/\zeta \leq p(z) \leq \zeta$, for all $z$ and some $\zeta \geq 1$.*

**Assumption 9.** *The kernel $k_h$ for the random cumulative incidence forest has bounded total variation. Specifically, there exists a constant $C$ such that the total variation of $k_h$ satisfies*

$$\int_{-\infty}^{\infty} |k_h'(x)| \, dx \leq C,$$

*where $k_h'(x)$ denotes the derivative of the kernel $k_h$.*

Below are assumptions for the proposed estimator for consistency. We assume bounding in Assumptions 10-11 to account for nonregularity of the value function. This is a necessary and standard assumption for non-regularity conditions.

**Assumption 10.** *Define the random function* $B(a, Z) = \phi_1^0(a, Z) - (1-\alpha_\phi)V_1^{0*}(Z)$, $a \in \mathcal{A}$. *Assume* $B(a, Z)$ *has continuous, bounded densities at 0, for each* $a \in \mathcal{A}$.

**Assumption 11.** $\sup_{a,Z} \left\| \begin{pmatrix} \phi_1^0(a, Z) \\ \phi_2^0(a, Z) \end{pmatrix} \right\|$ *is bounded by some* $\eta < \infty$.

## 4.1 Individualized Treatment Rule Estimator

For any fixed value of covariate $Z$, the difference in value functions between true optimal ITR and estimated optimal ITR based on Algorithm 2 is

$$\mathbb{V}(d^0_{\text{opt}}(Z)) - \mathbb{V}(\widehat{d}_{\text{opt}}(Z)) = \sum_{a\in\mathcal{A}} \mathbb{E}\left[ \begin{pmatrix} \phi_1^0(a, Z) \\ \phi_2^0(a, Z) \end{pmatrix} \cdot \Big( \mathbf{1}\{d^0_{\text{opt}}(Z) = a\} - \mathbf{1}\{\widehat{d}_{\text{opt}}(Z) = a\} \Big) \right] \quad (4)$$

where $\mathbb{V}(d(Z))$ is $\phi$ of a treatment regime $d(Z)$ if all individuals in a population follow the treatment regime. Showing Equation (4) converges in probability to 0 proves that $\widehat{d}_{\text{opt}}$ converges in probability to the true $d^0_{\text{opt}}$. The following theorem states that the individualized treatment regime estimator has a value consistent for the value of the optimal regime.

**Theorem 1.** *(Consistency of the individualized treatment regime estimator) Assume Assumptions 4-8 hold. Let* $\widehat{d}_{opt}$ *denote the optimal ITR estimator that is built following Algorithm 2 and Assumptions 10-11. Assume Assumptions 1-3 hold and that the random forests are built for each treatment arm. Then, for* $\phi^\mu = \begin{pmatrix} \phi_1^\mu \\ \phi_2^\mu \end{pmatrix}$,

$$\lim_{n\to\infty} \|\mathbb{V}(d^0_{opt}(Z)) - \mathbb{V}(\widehat{d}_{opt}(Z))\| \xrightarrow{p} 0$$

*where* $\|\cdot\|$ *is defined as the* $L_2$ *norm.*

## 4.2 Random Forests

Consistency of our proposed estimator depends on consistency of the forests. We show this in Theorems 2 and 3. For simplicity, let $h = (a, Z)$.

**Theorem 2** (Uniform consistency of a random survival forest)**.** *Suppose that Assumptions 4-8 hold. Let* $\widehat{S}(t \mid h)$ *be an estimator of the survival probability* $S(t \mid h)$ *that is uniformly consistent in both* $h \in \mathcal{H}$ *and* $t \in [0, \tau]$*. Then, the single-stage version of the random survival forest* $\widehat{S}(t \mid h)$ *in Cho et al. (2023) built based on* $\{H_i, \delta_i\}_{i=1}^n$ *is uniformly consistent. Specifically,*

$$\sup_{\substack{t\in(0,\tau] \\ h\in\mathcal{H}}} \left| \widehat{S}(t \mid h) - S(t \mid h) \right| \to 0$$

*in probability as* $n \to \infty$.

**Theorem 3** (Uniform consistency of a random cumulative incidence forest)**.** *Suppose that bounded kernels for the random cumulative incidence forest (RCIF) hold and have total bounded variation (Assumption 9). Let* $\widehat{F}_n(t \mid h)$ *be an estimator of the cause-j cumulative* *incidence function $F_j(t \mid h)$ that is uniformly consistent in both $h \in \mathcal{H}$ and $t \in [0, \tau]$. Then, the RCIF $\widehat{F}_n(t \mid h)$ built based on $\{H_i, \delta_i, \delta_{ji}\}_{i=1}^n$ is uniformly consistent. Specifically,*

$$\sup_{\substack{t\in(0,\tau] \\ h\in\mathcal{H}}} \left|\widehat{F}_n(t \mid h) - F_j(t \mid h)\right| \to 0$$

*in probability as $n \to \infty$.*

# 5 Simulations

We conducted an extensive simulation study to assess the finite sample performance of our proposed optimal ITR learning method. We examined two data generating mechanisms, an independent failure time simulation setting using exponential distributions and a dependent failure time simulation setting following Fine and Gray (1999). For both simulation settings, we simulated two causes of failure: priority cause 1 with failure time $T_1$ and cause 2 with failure time $T_2$. We generated $p$ covariates $Z_i$ from independent Uniform(0,1) distributions. We assumed two treatments, $A = 0, 1$. Let $\beta_1(A)$ and $\beta_2(A)$ denote the $p \times 1$ vectors of cause-specific coefficients dependent on treatment assignment $A$. The simulations varied by censoring level, treatment assignment mechanisms (independent of covariates vs conditional on covariates), training sample sizes, covariate dimensions, and (for Fine-Gray only) a low vs. high probability mass for the priority cause. Treatment was allocated using a Bernoulli trial with propensity $\pi(H) = \left[1 + \exp\left\{-Z_i^T\beta_\pi\right\}\right]^{-1}$, where the propensity coefficient is $\beta_\pi = (0, -\frac{1}{2}1_p)^T$ for the observational data setting and $0_{p+1}$ for the randomized trial setting.

We simulated 100 datasets and compared our method to dtrSurv which refers to the method in Cho et al. (2023), AIPWE which refers to the method in He et al. (2021), PMCR which refers to the method in Zhou et al. (2021), the zero-order model which is our proposed method but assumes no individual variation in treatment effects across patients, and the observed policy. We set the following forest parameters: $n_{\text{tree}} = 300$, $n_{\text{min}} = 5$ and $n_{\text{minevents}} = 2$. We chose $\alpha_\phi = 0.07$ for the truncated integration criterion $\phi(\cdot)$.We evaluated methods using a testing dataset with $N_{\text{eval}} = 1000$ subjects,assessing $\phi_1(a, Z)$ and $\phi_2(a, Z)$ across simulations. We calculated the OS and PC CIF value functions and averaged the criteria across all subjects per simulation.We note that our proposed method accommodates more than two treatments and causes.

## 5.1 Exponential Setting

In this simulation setting, we assume $T_1$ is independent from $T_2$, where $T_1 \sim \exp(\exp(Z_i^T\beta_1^A))$, $T_2 \sim \exp(\exp(Z_i^T\beta_2^A))$. Since $T_1 \perp T_2$, $T \sim \exp(\exp(Z_i^T\beta_1^A) + \exp(Z_i^T\beta_2^A))$. Let $\lambda_1^A = \exp(Z_i^T\beta_1^A)$ and $\lambda_2^A = \exp(Z_i^T\beta_2^A)$. Then for the truncated mean criterion, we have $\phi_1^\mu(t) = \int_0^\tau S(t)dt = \int_0^\tau \exp(-(\lambda_1^A + \lambda_2^A)t)dt$ and $\phi_2^\mu(t) = \int_0^\tau Pr(T \leq t, \epsilon = 1|Z_i)dt = \int_0^\tau \frac{\lambda_1^A}{\lambda_1^A+\lambda_2^A} - e^{-\lambda_2^A t} + \frac{\lambda_2^A}{\lambda_1^A+\lambda_2^A}\left(e^{-(\lambda_1^A+\lambda_2^A)t}\right) dt$. The derivations can be found in the Supplementary Material. Censoring is sampled from exponential distribution such that approximately 15% of the final dataset is censored for low censoring, or approximately 50% for high censoring.

In Figure 1, we plot the results from our simulation setting, where each dot represents a simulation iteration for each method. The left side shows the mean truncated OS and the right side shows the area under the cause 1 cumulative incidence curve.

Overall, the variance of the truncated mean criterion across simulations for our proposed method is generally smaller than those of other methods. The variance for simulations with $N = 700$ training data is smaller than that for $N = 300$, but there does not appear to be any obvious difference between the two treatment assigning mechanisms. The zero-order model only picks one treatment for the entire sample, thereby having a single regime across samples for the same setting. Thus, in the chance that the "worse" treatment is selected, it can be harmful to patients. We can see that compared to the zero-order model and observed policy, our proposed method has higher average survival. Compared to the proposed model, both dtrSurv and PMCR have lower average survival and higher average cumulative incidence. APIWE and the observed policy tend to have better CI but has a lower average survival, meaning they are more catered towards the priority cause, at the expense of OS.

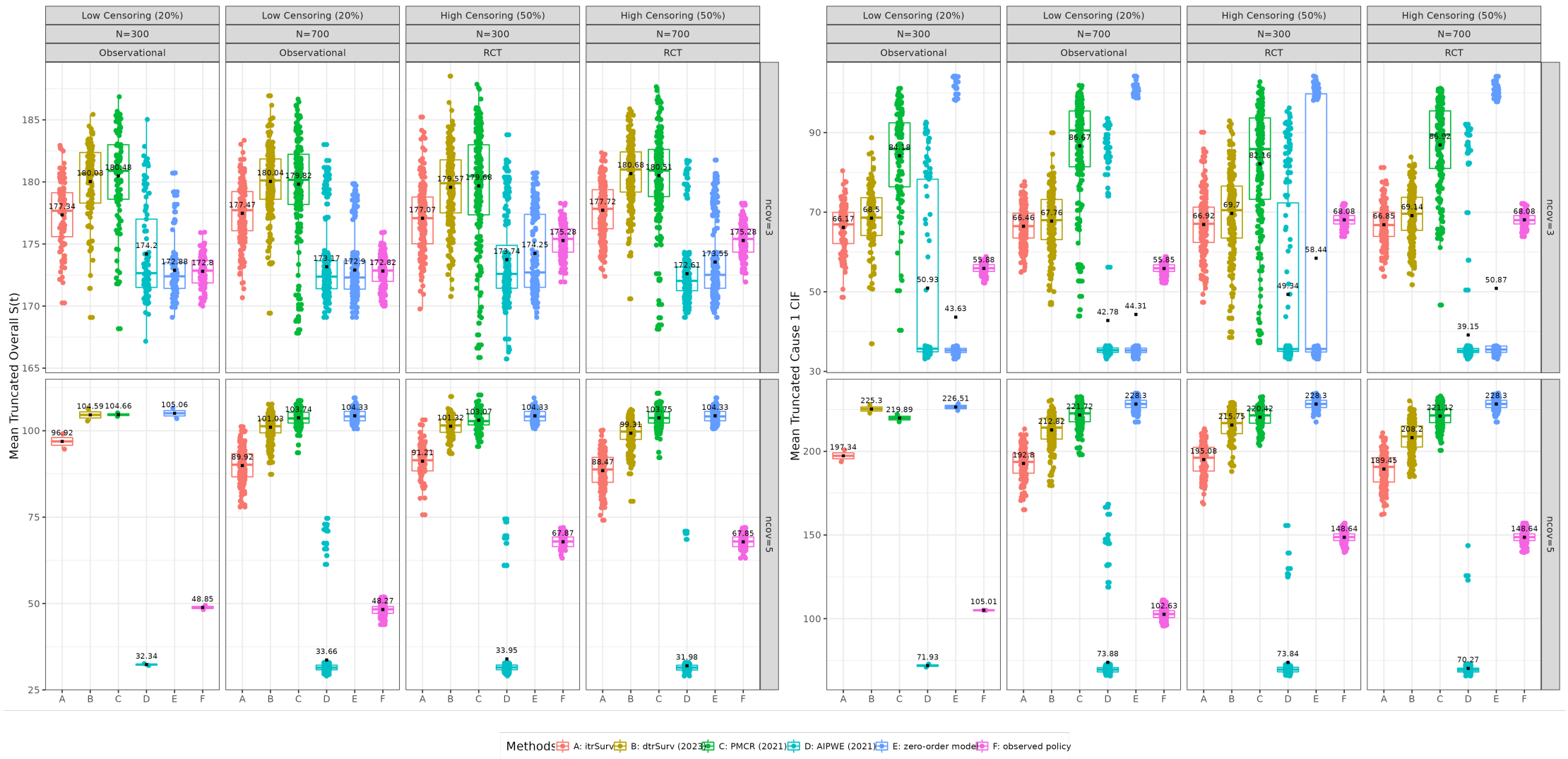

Figure 1: Results from exponential setting simulations. (Left) Estimated mean truncated OS time, and (right) the estimated integrated priority cause CIF under each estimated policy. Each dot represents one simulation replicate. The boxplots are the quartile summary of the points and the black dots are the average values of each method. The rows and columns of boxes represent a data generation scenario, study design and sample size.

## 5.2 Fine-Gray Setting

Following the simulation set-up in Fine and Gray (1999), we simulate two competing risks for $i = 1, \ldots, N$, where the CIF for failures from cause 1 ($\epsilon = 1$) are given by a unit exponential mixture with mass $1-p$ at $\infty$ when covariate values are 0, and use proportional subdistribution hazards model to obtain CIF for nonzero covariate values:

$$\Pr(T_i \leq t, \epsilon = 1 | Z_i) = 1 - [1 - p\{1 - exp(-t)\}]^{\exp(Z_i^T \beta_1^A)}.$$

More details are provided in the supplementary material. We first examine the mean truncated restricted OS time ($\phi_1^\mu = \int_0^\tau S(t)dt$), then the mean truncated restricted PC (cause 1) CIF ($\phi_2^\mu = \int_0^\tau \Pr(T \leq t, \epsilon = 1)dt$). The full simulation results are displayed in Figure 2, where each dot represents a simulation iteration for each method, and the average across simulations is presented. A higher truncated mean survival time (left) and lower area under CIF for priority cause (right) is ideal.

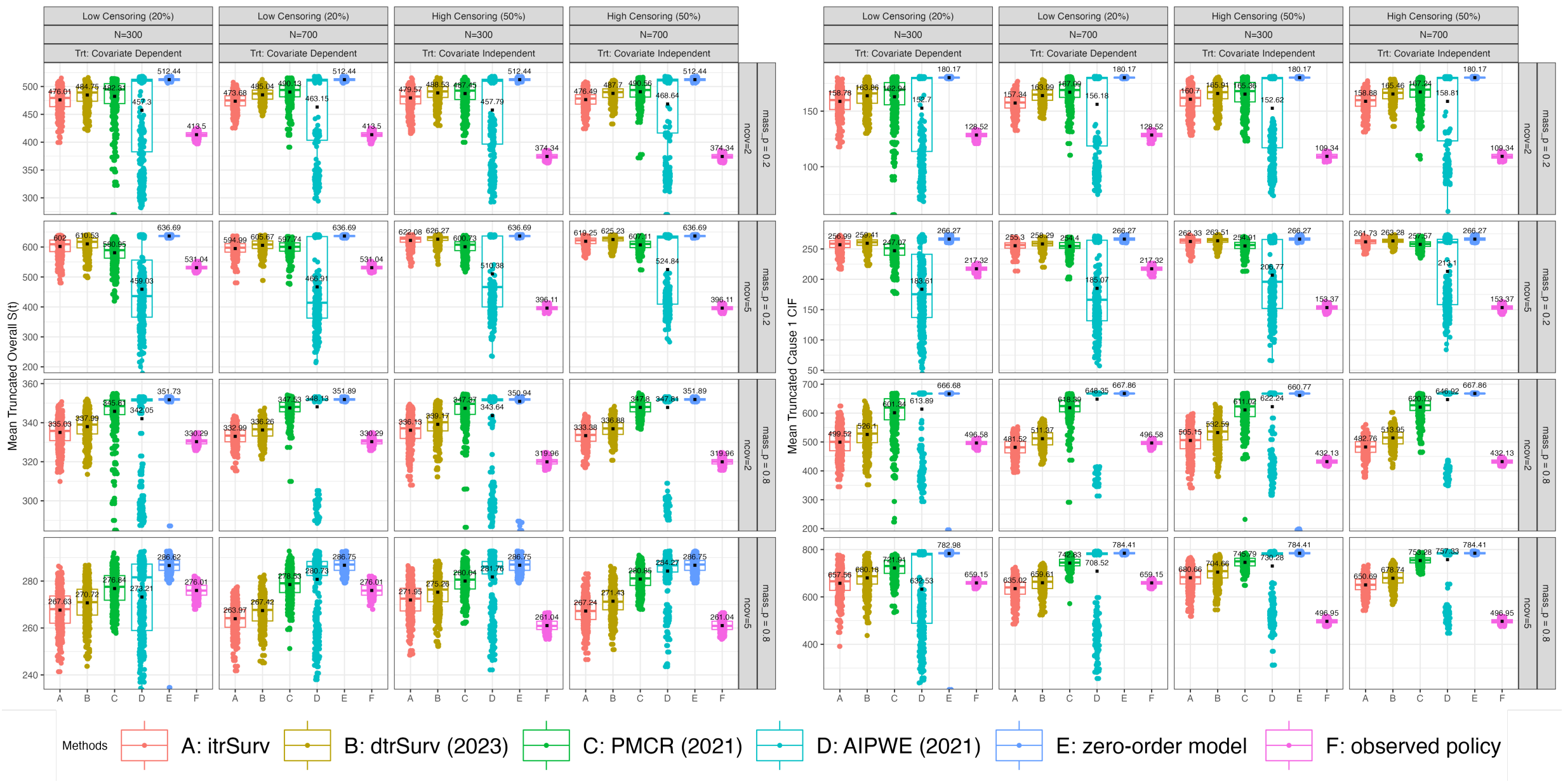

Figure 2: Results from Fine-Gray setting simulations. (Left) Estimated truncated mean overall survival time, and (right) the estimated integrated priority cause cumulative incidence function under each estimated policy. Each dot represents one simulation replicate. The boxplots are the quartile summary of the points and the black dots are the average values of each method. The rows and columns of boxes represent a data generation scenario, study design and sample size.

When $p_{\text{mass}} = 0.2$, our proposed method generally has the same mean times for OS (albeit slightly lower) and slightly lower mean times for CI. We note that while AIPWE and the observed policy sometimes have a lower CIF average (e.g., 5 covariates and $p_{\text{mass}} = 0.2$), their mean survival time average is also much lower. We also note that in the second row, our method has a mean lower than dtrSurv but higher than AIPWE and PMCR; and the opposite is true for cumulative incidence (lower compared to dtrSurv, but higher compared to AIPWE and PMCR). This is what we would expect, since dtrSurv determines optimal ITRs based on OS, while AIPWE and PMCR determine based on cumulative incidence. When $p_{\text{mass}} = 0.8$, where PC has strong influence on the data, the OS for all methods are similar with low variance across simulations (see Figure 3 for fixed y-axis scales). However, when comparing the PC cumulative incidence, our model produces the lowest times on average, with the exception of the observed policy (which also has very low OS on average).

Figure 3 displays just the setting of low censoring, high sample size, and 5 covariates where the treatment assignment did not depend on covariates and $p_{\text{mass}} = 0.8$ from Figure 2. We can see that our method performs well, with similar survival times to other methods while producing lower mean truncated cumulative incidence for the PC times on average across simulations.

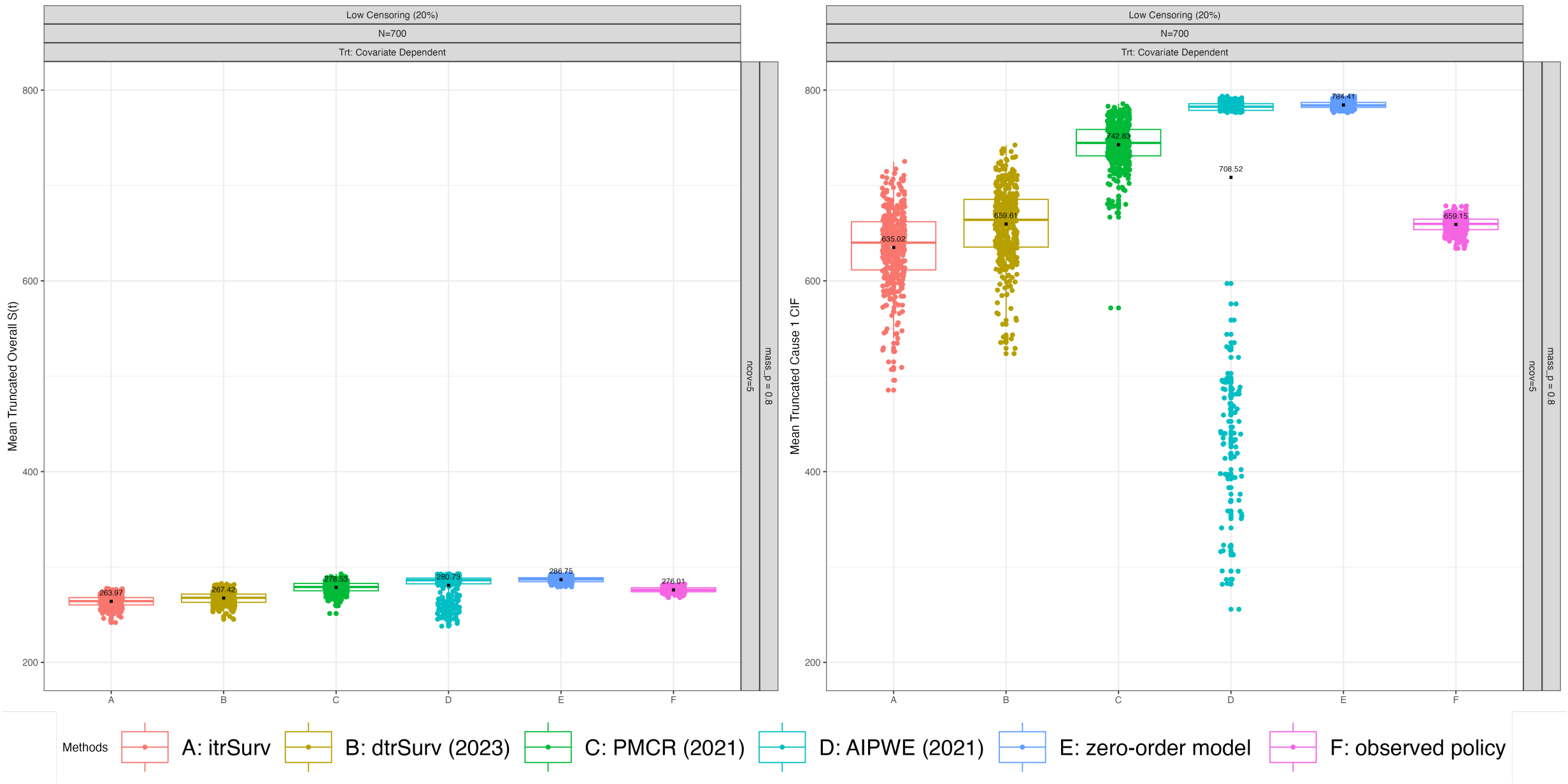

Figure 3: Fine-Gray setting for low censoring, high sample size, high covariates, high PC probability mass for covariate-independent treatment assignment mechanism. (Left) Estimated truncated mean overall survival time, and (right) the estimated integrated priority cause cumulative incidence function under each estimated policy. Each dot represents one simulation replicate. The boxplots are the quartile summary of the points, and the black dots are the average values of each method.

# 6 Analyses of Peripheral Artery Disease Cohort

We applied our proposed method to a cohort of patients treated for CLTI in a single academic medical center. The dataset was derived from the Carolina Data Warehouse for Health, which is housed within the North Carolina Translational and Clinical Sciences Institute (NC TraCS) and includes electronic health record data derived from care provided at the University of North Carolina at Chapel Hill Medical Center. The cohort included patients that met the hemodynamic and symptomatic diagnostic criteria for CLTI and underwent an open surgical or endovascular revascularization to restore blood flow to the lower limbs between April 2013 and October 2017. Details for this cohort were described previously (McGinigle et al., 2021).

CLTI predominantly affects older adults, who have a history of smoking or other risk factors for developing arterial occlusive disease like hypertension and elevated cholesterol. Five-year mortality after a diagnosis of CLTI is 60% (Criqui et al., 1992; van Haelst et al., 2018; Baubeta Fridh et al., 2017). Poor systemic cardiovascular health increases the risk of heart attacks, stroke, and premature death, the severe arterial insufficiency in the arteries of the legs puts patients with CLTI patients at risk for non-healing foot ulcers, gangrene, or foot infections, all of which further elevate the risk of limb loss (Mills et al., 2014; Browder et al., 2022; Crowner et al., 2022). However, despite multiple treatment options, approximately half of patients with CLTI will experience limb loss or death within two years following surgical intervention (Farber et al., 2022). Our goal is to derive the optimal ITR for patients undergoing the two surgical treatments, endovascular therapy and open bypass surgery, by prioritizing OS while minimizing the risk of death, with major amputation before death considered as a competing risk.

Patients with more than one limb affected by disease were excluded from the study. Among those who experienced major amputation, we only included patients who had the amputation on the limb that underwent the index revascularization procedure. For instance, patients who had major amputation on the left limb but received open surgical revascularization on the right limb were excluded. Given that the decision to perform an open bypass is largely based on surgeon and patient preference, which may be influenced by the invasive nature of the procedure, we classified patients who initially underwent endovascular surgery (such as angioplasty or stent) but subsequently received an open bypass within a week as part of the open bypass treatment group. Our final analytic dataset had $N = 303$, with 77.9% of patients in the endovascular treatment group and 22.1% in the open bypass group. We analyzed 1-year follow-up data such that $\tau = 365$ days, and any observed events after $\tau$ were considered censored with $X_i = \tau$. The tailoring variables were age at treatment (in years), race, ethnicity, gender, severity of clinical presentation (limb-based Rutherford class), degree of ischemia, severity of foot wound anatomic location of arterial occlusive disease, previous revascularization history, and presences of clinical significant comorbidities including diabetes, hypertension, hyperlipidemia, renal disease, coronary artery disease, myocardial infarction, uncomplicated or complicated congestive heart failure, cerebrovascular accident, obesity, smoking, chronic obstructive pulmonary disease, coagulopathy, anemia, dementia, venous insufficiency, venous thromboembolism (VTE), and unintentional weight loss. For more details, refer to the Supplementary Material.

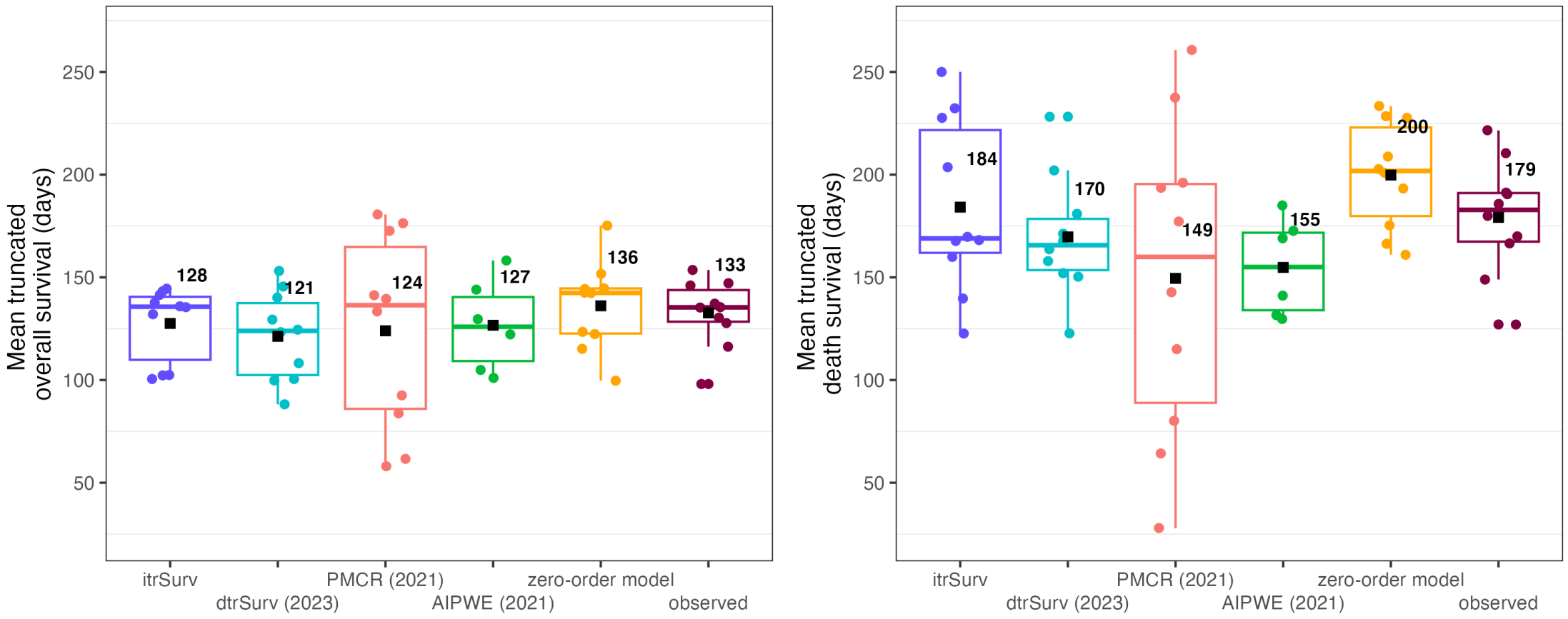

Figure 4: (Left) Estimated truncated mean survival time for preventing failure from all causes (major amputation before death or death) and (right) estimated truncated mean survival time for preventing mortality of the individualized treatment regime estimators. Each dot represents one cross-validation replicate. The black square dots are the average values of each method.

Our results compared to the same methods in the simulation studies, the zero-order model, and the observed policy are presented in Figure 4. We illustrated the truncated mean failure time for both endpoints, where failure on the left is failure from any cause, and failure on the right is failure from death, the priority cause. AIPWE was unable to perform with the sample size less than 500 for all CV iterations. Thus, we did not include all $K_{\text{CV}} = 10$ folds AIPWE in our results. From Figure 4, our method performs well compared to other existing precision medicine methods. We note that the zero-order model version of our method may outperform the precision medicine version of our method. However, this implies that our proposed two-phase method works well, but heterogeneity might not be enough in this dataset to distinguish between patients. We have already show via simulations that heterogeneity is important to consider. The observed policy average also performed slightly better than our method average for mean truncated OS time. This suggested that the surgeon treatment recommendation is good for the patients, as expected from the experts in the field. Because our method also considers minimizing cumulative incidence of death, the average across folds for mean truncated death failure time is still higher than the observed policy. However, limitations in this analysis include an unbalanced number in the treatment groups and high censoring rate (70%).

# 7 Discussion

We proposed a novel precision medicine approach utilizing multi-utility value functions for survival outcomes in the presence of competing risks to estimate optimal individualized treatment regimes. Our main contributions are: 1) estimating optimal ITRs based on OS and priority cause cumulative incidence simultaneously, 2) prioritizing maximum OS over minimum priority cause risk, 3) introducing a multi-utility value function, and 4) developing random survival and cumulative forests that output individualized CIF curves for single stage treatment allocation. Under standard assumptions and mild regularity conditions, we established uniform consistency of the value function of our optimal ITR estimator, as well as for the proposed survival and cumulative incidence random forests.

Our work builds on the work of Cho et al. (2023), which is designed for finite decision points for survival outcomes and do not consider cause-specific events. Additionally, our work extends utility functions for multiple outcomes (Butler et al., 2018; Jiang et al., 2021; Zitovsky et al., 2023) by presenting a multi-utility approach. Our work is closest to Jiang et al. (2021), but they used multiple utility functions to construct a composite of equally weighted outcomes, while we implement it directly, incorporating hierarchical structure through a two-step procedure. With respect to the existing CR PM methods (Friedenreich et al., 2016; Yavuz et al., 2018; Chen et al., 2020; Morzywołek et al., 2022), our work most closely relates to Zhou et al. (2021), who minimized side effects in a CR setting by estimating the optimal ITR at a specific time point, with side effects constrained below a set threshold. While our method is similar, we incorporate a pre-specified tolerance to determine whether the CR is applied. Additionally, Zhou et al. (2021) only considered linear decision rules whereas our approach uses flexible forest-based ensemble methods to capture complex data patterns. We also compared to He et al. (2021), who's approach is doubly robust, while our nonparametric method entirely avoids the risk of model misspecification by not relying on parametric assumptions. He et al. (2021) only considers binary treatments. In contrast, our method accommodates a finite number of treatments.

Our proposed method has a few limitations. It assumes that one cause is more important than others, which could affect its applicability in cases where the relative importance of causes is unclear or varies. Additionally, our method does not include a tie-breaker mechanism, meaning optimal ITRs may not be unique. Finally, the user determines the pre-specified tolerance for our method, making it dependent on their judgment and expertise. In this manuscript, we provide a meaningful advancement for decision-making in complex scenarios through multi-utility optimization. Future research may include incorporating tiebreakers for scenarios with multiple optimal ITRs or extending to multi-stage treatment settings. Such investigations would provide valuable insights into the method's versatility.

# 8 Acknowledgment

The authors thank Sydney Browder for her help cleaning the cohort study data, which was from the 2017 UNC Center for Healthcare Innovation pilot grant.

# 9 Supplementary Material

The Supplementary Material contains more information on the random forests, all theorem proofs, and more details on the simulation studies and PAD analysis. The R package is available at https://github.com/cwzhou/itrSurv. All R code used in the simulations and application are available at https://github.com/cwzhou/Analyses.

# Supplementary Material for "Optimal individualized treatment regimes for survival data with competing risks"

# A Notation Table

See Table S1 for notation summary.

# B Proposed Random Forests

## B.1 Brief Review of Tree-based Methods

A random forest is an ensemble of trees, which are generated by recursively partitioning the data into two daughter leaves, or nodes. Established tree-based approaches include Classification and Regression Trees (CART) (Breiman et al., 1984), CART extensions (Breiman et al., 1984), and Extremely Randomized Trees (ERT) (Geurts et al., 2006). A random survival forest is an extension of the random forest that accounts for censoring by allowing censored outcomes as inputs and providing covariate-conditional survival probability estimates as outputs (Ishwaran et al., 2008). This is a nonparametric survival regression estimator that provides weighted Kaplan-Meier curves for each point in the feature space with weights given by adaptive kernel estimates. Because of the presence of censoring, it is not feasible to use a simple mean for splitting rules. Thus, the log-rank test is used (Peto and Peto, 1972; Gehan, 1965; Tarone and Ware, 1977). At each terminal node of a survival tree, the survival probability is estimated by the Kaplan-Meier estimator. A random survival forest is obtained by averaging the survival trees.

Table S1: Summary of notation

| | |
|---|---|
| **Basic notation** | |
| $\tau$ | Study length |
| $n$ | Number of training examples |
| $\epsilon$ | Event cause |
| $i$ | Subject index. 1, 2, ..., $n$ |
| $j$ | Cause index. 1, 2, ..., $J$ |
| $k$ | Group or Node index. 1, 2, ..., $K$ |
| **Predictors** | |
| $A$ | Action |
| $Z$ | Covariate information |
| $H$ | History $(a, Z)$ |
| $d(Z)$ | Decision rule given $Z$ |
| **Overall outcomes** | |
| $T$ | Failure time due to any cause. $T = \min(T_1, T_2, \ldots, T_J)$ |
| $C$ | Overall censoring time |
| $X$ | Observed length. $X = T \wedge C$ |
| $\delta$ | Overall Censoring indicator. $\delta = 1(T \leq C)$ |
| **Cause-specific outcomes** | |
| $T_j$ | Failure time due to cause $j$ |
| $\delta_j$ | Cause $j$ Censoring indicator. $\delta_j = 1(T \leq C, T = T_j)$ |
| **Theoretical settings** | |
| $L_S, L_G$ | Lipschitz constants of Assumption 7 |
| $n_{\min}$ | Minimum terminal node size in Assumptions 5 |
| $\beta$ | Rate of minimum terminal node size in Assumption 5 |
| $\alpha$ | Regular split constant in Assumption 6 |
| $\zeta$ | Covariate density bound in Assumptions 8 |
| $c_1$ | At-risk probability bound at $\tau$ in Assumption 7 |
| **Survival and cumulative incidence functions** | |
| $S$ | Overall failure survival function |
| $F_j$ | Cause $j$ cumulative incidence function. $F_j(t) = \Pr(T \leq t, \epsilon = j)$ |
| $G$ | Overall or generic censoring survival function |
| **Values** | |
| $\phi_1^\mu(S)$ | Mean truncated survival time. $\phi^\mu(S) = E[S \wedge \tau]$ |
| $\phi_2^\mu(F_j)$ | Mean truncated cumulative incidence time. $\phi^\mu(F_j) = E[F_j \wedge \tau]$ |

## B.2 Proposed Forests

For our forest models, let $n_{tree}$ denote the number of trees. We assume that $n_{\text{tree}} \geq 1$. A larger $n_{\text{tree}}$ is preferred in practice. Nevertheless, all theoretical results presented in this paper remain valid for all $n_{\text{tree}} \geq 1$.

## Random Survival Forest

We adapt the forest algorithm from Cho et al. (2023) to single stage treatment settings. We provide a high-level algorithm of the forest in Algorithm S1. Briefly, we use the Kaplan-Meier estimator to estimate the marginal survival probability of a node (or leaf) $\mathcal{L}$, where each individual's failure time information is given by $\{(X_i, \delta_i)\}_{i=1}^{n_{\mathcal{L}}}$ where $n_{\mathcal{L}}$ is the size of node $\mathcal{L}$, $X_i$ is the observed failure time, and $\delta_i$ is the event indicator for any cause:

$$\widehat{S}(t \mid \mathcal{L}) = \prod_{s \geq 0} \left(1 + \frac{\sum_{i \in \mathcal{L}} dN_i(s)}{\sum_{i \in \mathcal{L}} Y_i(s)}\right), \tag{S1}$$

where $N_i(t) = I(X_i \leq t, \delta_i = 1)$ and $Y_i(t) = I(X_i \geq t)$. The Kaplan-Meier estimator is used in both tree partitioning and terminal node survival probability estimation.

When the optimization criterion is the truncated mean survival time ($\phi = \phi^\mu$), the forest algorithm splits a node into two candidate daughter nodes, $\mathcal{L}_1$ and $\mathcal{L}_2$ by using the mean difference test to maximize the difference in truncated mean survival times:

$$MD(\mathcal{L}_1, \mathcal{L}_2) = \left| \int_{t>0} (t \wedge \tau) d\widehat{S}(t \mid \mathcal{L}_1) - \int_{t>0} (t \wedge \tau) d\widehat{S}(t \mid \mathcal{L}_2) \right|. \tag{S2}$$

This splitting rule identifies the best split as $(\mathcal{L}_1, \mathcal{L}_2) = \arg\max_{\mathcal{L}_1, \mathcal{L}_2} MD(\mathcal{L}_1, \mathcal{L}_2)$.

The recursive partitioning process continues as long as the size of each node remains above the threshold $n_{\min}$. The survival probability estimate of the terminal node $\mathcal{L}$ of the $w$-th ($w = 1, 2, \ldots, n_{\text{tree}}$) survival tree is given by the Kaplan-Meier estimator $\widehat{S}_w(t \mid \mathcal{L})$. Letting $h = (a, Z)$, the random survival forest estimate is then given by

$$\widehat{S}(t \mid h) = \frac{1}{n_{\text{tree}}} \sum_{w=1}^{n_{\text{tree}}} \sum_{l=1}^{\#\{\mathcal{L}_{\{w\},l}:l\}} 1(h \in \mathcal{L}_{\{w\},l}) \widehat{S}(t \mid \mathcal{L}_{\{w\},l}),$$

where $\#\{\mathcal{L}_{\{w\},l} : l\}$ is the number of terminal nodes of the $w$-th tree.

---

**Algorithm S1** The random survival forest estimator

**Input:** $\{(X_i, \delta_i, Z_i, A_i)\}_{i=1}^n$

**Output:** $\widehat{S}(\cdot|Z, A = a)$

Result: random survival forest estimate $\widehat{S}(\cdot|Z, A)$

Parameters: the number of trees $n_{\text{tree}}$, the minimum terminal node size $n_{\min}$

For treatment arm $a \in \mathcal{A}$ **do**

For tree $w = 1, 2, \ldots, n_{\text{tree}}$ **do**

Partition the feature space with the input data $\{(I(t \leq X_i), \delta_{ji}, Z_i)\}_{i:a_i=a}$ via MD test (S2) until node size $< 2n_{\min}$

Obtain the terminal node survival probability estimate $\widehat{S}\{w\}(t|a, Z)$ via Kaplan-Meier estimator (S1)

Obtain the forest estimator $\widehat{S}(\cdot|Z, A = a) = \frac{1}{n_{\text{tree}}} \sum_{w=1}^{n_{\text{tree}}} \widehat{S}_{\{w\}}(t|a, Z)$

---

# C Proof of Theorem 1

*Proof of Theorem 1.* Note that

$$\begin{aligned}
\left\|\mathbb{V}(d^0_{\text{opt}}(Z)) - \mathbb{V}(\widehat{d}_{\text{opt}}(Z))\right\| &= \left\|\sum_{a\in\mathcal{A}} \mathbb{E}_Z\left[\begin{pmatrix}\phi^0_1(a,Z)\\ \phi^0_2(a,Z)\end{pmatrix} \cdot \Big(\mathbf{1}\{d^0_{\text{opt}}(Z) = a\} - \mathbf{1}\{\widehat{d}_{\text{opt}}(Z) = a\}\Big)\right]\right\| \\
&\leq \sum_{a\in\mathcal{A}} \left\|\mathbb{E}_Z\left[\begin{pmatrix}\phi^0_1(a,Z)\\ \phi^0_2(a,Z)\end{pmatrix} \cdot \Big(\mathbf{1}\{d^0_{\text{opt}}(Z) = a\} - \mathbf{1}\{\widehat{d}_{\text{opt}}(Z) = a\}\Big)\right]\right\| \\
&\leq \sum_{a\in\mathcal{A}} \mathbb{E}_Z\left\|\left[\begin{pmatrix}\phi^0_1(a,Z)\\ \phi^0_2(a,Z)\end{pmatrix} \cdot \Big(\mathbf{1}\{d^0_{\text{opt}}(Z) = a\} - \mathbf{1}\{\widehat{d}_{\text{opt}}(Z) = a\}\Big)\right]\right\| \\
&\leq \sum_{a\in\mathcal{A}} \mathbb{E}_Z\left[\left\|\begin{pmatrix}\phi^0_1(a,Z)\\ \phi^0_2(a,Z)\end{pmatrix}\right\| \cdot \left|\mathbf{1}\{d^0_{\text{opt}}(Z) = a\} - \mathbf{1}\{\widehat{d}_{\text{opt}}(Z) = a\}\right|\right] \\
&= \mathbb{E}_Z\left[\sum_{a\in\mathcal{A}} \left(\mathbf{1}\{d^0_{\text{opt}}(Z) \neq \widehat{d}_{\text{opt}}(Z)\} \left\|\begin{pmatrix}\phi^0_1(a,Z)\\ \phi^0_2(a,Z)\end{pmatrix}\right\|\right)\right] \\
&\leq \mathbb{E}_Z\left(\mathbf{1}\{d^0_{\text{opt}}(Z) \neq \widehat{d}_{\text{opt}}(Z)\}\right) \sup_{a,Z} \left\|\begin{pmatrix}\phi^0_1(a,Z)\\ \phi^0_2(a,Z)\end{pmatrix}\right\| \\
&= \Pr(d^0_{\text{opt}}(Z) \neq \widehat{d}_{\text{opt}}(Z)|\widehat{d}_{\text{opt}}) \sup_{a,Z} \left\|\begin{pmatrix}\phi^0_1(a,Z)\\ \phi^0_2(a,Z)\end{pmatrix}\right\| \\
&\leq \eta \Pr(d^0_{\text{opt}}(Z) \neq \widehat{d}_{\text{opt}}(Z)|\widehat{d}_{\text{opt}}).
\end{aligned}$$

The third to last line holds because 1) only one term in the summation will have a non-zero contribution and 2) the supremum represents the "worst case scenario" over all fixed combinations of $a$ and $z$ and thus does not depend on the random variables. The last line holds because we assumed $\sup_{a,Z}\left\|\begin{pmatrix}\phi^0_1(a,Z)\\ \phi^0_2(a,Z)\end{pmatrix}\right\|$ is bounded by $\eta$ via Assumption 11. Thus, the norm of this distance, $\mathbb{V}(d^0_{\text{opt}})-\mathbb{V}(\widehat{d}_{\text{opt}})$, goes to 0 in probability if $\Pr(d^0_{\text{opt}}(Z) \neq \widehat{d}_{\text{opt}}(Z)|\hat{d}_{\text{opt}}) \xrightarrow{p} 0$.

We complete the proof by first showing

$$\sum_{a\in\mathcal{A}} \left| I\{\phi^0_1(a,Z) - (1-\alpha_\phi)V^{0*}_1(Z) \geq 0\} - I\{\widehat{\phi}_1(a,Z) - (1-\alpha_\phi)\widehat{V}^*_1(Z) \geq 0\} \right| \to 0. \quad \text{(S3)}$$

The proof of (S3) will be presented shortly. Now, (S3) implies

$$\underset{a\in\mathcal{A}:\widehat{\phi}_1(a,Z)\geq(1-\alpha_\phi)\widehat{V}^*_1(Z)}{\arg\max} \widehat{\phi}_2(a,Z) \xrightarrow{p} \underset{a\in\mathcal{A}:\phi^0_1(a,Z)\geq(1-\alpha_\phi)V^{0*}_1(Z)}{\arg\max} \widehat{\phi}_2(a,Z). \quad \text{(S4)}$$

Theorem 3 in our paper states that RCIFs are consistent. Thus, by the Continuous Mapping Theorem, $\widehat{\phi}_2(a,Z) \xrightarrow{p} \phi^0_2(a,Z)$. Then,

$$\sup_{a\in\mathcal{A}} \mathbb{E} \left| \widehat{\phi}_2(a,Z) - \phi^0_2(a,Z) \right| \xrightarrow{p} 0, \quad \text{(S5)}$$

which implies that the right hand side of (S4) converges to
$\arg\max_{a\in\mathcal{A}:\phi^0_1(a,Z)\geq(1-\alpha_\phi)V^{0*}_1(Z)} \phi^0_2(a,Z)$. Thus, since $Z$ is random,

$$\Pr(d^0_{\text{opt}}(Z) \neq \widehat{d}_{\text{opt}}(Z)|\widehat{d}_{\text{opt}}) \xrightarrow{p} 0.$$

This completes the proof of Theorem 1. □

*Proof of (S3).* Let $A_0 = \phi^0_1(a,Z) - (1-\alpha_\phi)V^{0*}_1(Z)$ and $\widehat{A}_n = \widehat{\phi}_1(a,Z) - (1-\alpha_\phi)\widehat{V}^*_1(Z)$. These are functions of random variable $Z$. Then, we can re-write (S3) as

$$\sum_{a\in\mathcal{A}} \left| I\{A_0 \geq 0\} - I\{\widehat{A}_n \geq 0\} \right| \to 0.$$

$$\begin{aligned} \left| I\{\widehat{A}_n \geq 0\} - I\{A_0 \geq 0\} \right| &= I\{\widehat{A}_n < 0 \leq A_0\} + I\{A_0 < 0 \leq \widehat{A}_n\} \\ &= I\{\widehat{A}_n - A_0 < -A_0 \leq 0\} + I\{0 < -A_0 \leq \widehat{A}_n - A_0\} \\ &= I\{0 \leq A_0 < -(\widehat{A}_n - A_0)\} + I\{-(\widehat{A}_n - A_0) \leq A_0 < 0\}. \end{aligned} \quad \text{(S6)}$$

In (S6), the first equality holds because these are the only possible indicators such that their sum is equal to 1. We get the 2nd equality by subtracting $A_0$ on both sides within the indicator functions. The third equality is due to multiplying by $-1$ and flipping the inequalities within the indicators.

Since we know $\widehat{\phi}_1(a,Z) \xrightarrow{p} \phi^0_1(a,Z)$ and $\widehat{V}^*_1(Z) \xrightarrow{p} V^{0*}_1(Z)$ as shown in Cho et al. (2023); and $\widehat{A}_n$ and $A_0$ are linear combinations of $\widehat{\phi}_1(a,Z)$ and $\widehat{V}^*_1(Z)$, and $\phi^0_1(a,Z)$ and $V^{0*}_1(Z)$, respectively, then $\widehat{A}_n \xrightarrow{p} A_0$ because the linear combination of consistent estimators is consistent. This means

$$\widehat{A}_n - A_0 \xrightarrow{p} 0. \quad \text{(S7)}$$

By definition of convergence in probability, (S7) implies $\exists\epsilon_n : Pr(|\widehat{A}_n - A_0| > \epsilon_n) \to 0$ as $\epsilon_n \to 0$ for a fixed sequence $\epsilon_n$. Then, using Assumption 10 and (S7), we can bound (S6). Starting with the first term of (S6),

$$I\{0 \leq A_0 < -(\widehat{A}_n - A_0), |\widehat{A}_n - A_0| \leq \epsilon_n\} + I\{0 \leq A_0 < -(\widehat{A}_n - A_0), |\widehat{A}_n - A_0| > \epsilon_n\} \leq$$
$$I\{0 \leq A_0 < \epsilon_n\} + I\{|\widehat{A}_n - A_0| > \epsilon_n\},$$

where $I\{0 \leq A_0 < \epsilon_n\} \to 0$ by Assumption 10 (bounded density assumption) and $I\{|\widehat{A}_n - A_0| > \epsilon_n\} \to 0$ by (S7) (and definition of consistency). Specifically, assuming Assumption 10, $Pr(0 \leq A_0 \leq \epsilon_n) = \int_0^{\epsilon_n} f_{A_0}(a_0) da_0 \approx f_{a_0}(0)\epsilon_n \leq b\epsilon_n \to 0$ as $\epsilon_n \to 0$ for some bound $b$. The second term can be shown to be bounded similarly.

$$
\begin{aligned}
&I\{-(\widehat{A}_n - A_0) \leq A_0 < 0\} = \\
&I\{-(\widehat{A}_n - A_0) \leq A_0 < 0, |\widehat{A}_n - A_0| \leq \epsilon_n\} + I\{-(\widehat{A}_n - A_0) \leq A_0 < 0, |\widehat{A}_n - A_0| > \epsilon_n\} \leq \\
&I(-\epsilon_n \leq A_0 < 0) + I(|\widehat{A}_n - A_0| > \epsilon_n).
\end{aligned}
$$

The first term for the inequality is because $-(\widehat{A}_n - A_0) < 0$, $\widehat{A}_n - A_0$ must be positive; thus, $-\epsilon_n \leq \widehat{A}_n - A_0 \leq \epsilon_n$. The second term is because the indicator is a subset.

Because of Equation (S7), $Pr(|\widehat{A}_n - A_0| > \epsilon_n) \to 0$, and $Pr(-\epsilon_n \leq A_0 < 0) \to 0$ due to Assumption 10. Specifically, $I(-\epsilon_n \leq A_0 < 0) \implies Pr(-\epsilon_n \leq A_0 < 0) \approx \epsilon_n f_{a_0}(0) < \epsilon_n b \to 0$ as $\epsilon_n \to 0$. This completes the proof of (S3). □

# D Proof of Theorem 2

*Proof of Theorem 2.* It is sufficient to establish the consistency results for a single tree in order to prove Theorem 2. By nature of the forest, which is the average of multiple trees where the randomization process of the trees is independent of the data generative law, the supremum error of averages is less than the average of supremum errors for the uniform consistency, i.e., for $\hat{S} = \frac{1}{n_{\text{tree}}} \sum_{w=1}^{n_{\text{tree}}} \hat{S}^{(w)}$,

$$
\sup_{t,h} \left| \hat{S}(t_j|h) - S(t_j|h) \right| \leq \frac{1}{n_{\text{tree}}} \sum_{w=1}^{n_{\text{tree}}} \sup_{t,h} \left| \hat{S}^{(w)}(t_j|h) - S(t_j|h) \right|.
$$

Hence, we will prove prove Theorem 1 for just a single tree $\widehat{S}_{\{w\}}$, which is built based on $\{H_i, \delta_i\}_{i=1}^n$. For ease of notation, we drop the tree index $w$ throughout the proof.

Using a modified version of Lemma 2.10 from Kosorok (2008), let $\Psi : \Theta \to L$ and $\Psi_n : \Theta \to L$ be maps between two normed spaces, where $\Theta$ is a parameter space, $L$ is some normed space, $\|\cdot\|_L$ denotes the uniform norm, $\Psi$ is a fixed map, and $\Psi_n$ is a data-dependent map. This lemma is modified to allow $\Theta_n$ to be data-dependent. Let $\Theta_n$ be the data-dependent survival function space, where survival functions are piecewise constant over a hyper-rectangle, represented by a rectangular kernel $k_h$ in the feature space so that $S(t \mid k_h) = S(t \mid h)$ for all $h \in \mathcal{H}$.

**Lemma 1** (Consistency of Z-estimators)**.** *Let $\Psi(S_0) = 0$ for some $S_0 \in \Theta$, and assume $\|\Psi(S_n)\|_L \to 0$ implies $\|S_n - S_0\|_\infty \to 0$ for any sequence $\{S_n\}_{n\in\mathbb{N}} \subset \Theta_n \to \Theta$ as $n \to \infty$. Then, if $\left\|\Psi_n(\hat{S}_n)\right\|_L \to 0$ in probability for some sequence of estimators $\{\hat{S}_n\}_{n\in\mathbb{N}} \subset \Theta_n$ and $\sup_{S\in\Theta_n} \|\Psi_n(S) - \Psi(S)\|_L \to 0$ in probability, $\left\|\hat{S}_n - S_0\right\|_\infty \to 0$ in probability, as $n \to \infty$.*

*Proof of Lemma 1.* Suppose a sequence $\hat{S}_n \in \Theta_n$ satisfies that $\left\|\Psi_n(\hat{S}_n)\right\|_L \to 0$ and

$\sup_{S\in\Theta_n} \|\Psi_n(S) - \Psi(S)\|_L \to 0$ both in probability as $n \to \infty$. Then we have

$$\left\|\Psi(\hat{S}_n)\right\|_L \leq \left\|\Psi(\hat{S}_n) - \Psi_n(\hat{S}_n)\right\|_L + \left\|\Psi_n(\hat{S}_n)\right\|_L$$
$$\leq \sup_{S\in\Theta_n} \|\Psi(S) - \Psi_n(S)\|_L + \left\|\Psi_n(\hat{S}_n)\right\|_L \to 0,$$

in probability as $n \to \infty$. Thus, by the assumption, $\left\|\hat{S}_n - S_0\right\|_\infty \to 0$. □

Let $\Theta$ be the space of all covariate-conditional survival functions. Define a normed space $L = D_{[-1,1]}\{[0,\tau] \times \mathbb{R}^d\}$, where $D_A B$ is the space of all right-continuous left-limits functions with range $A$ and support $B$. We use $S_0$ to denote the true survival function.

We define the following estimating function and equations:

$$\psi_{S,t} = I(X > t) + (1-\delta)I(X \leq t)\frac{S(t \mid H)}{S(X \mid H)} - S(t \mid H), \tag{S8}$$

$$\Psi(S) = \frac{P\{\psi_{S,t}\delta_h\}}{P\{\delta_h\}} \equiv P_{\cdot|h}\psi_{S,t}, \tag{S9}$$

$$\Psi_n(S) = \frac{\mathbb{P}_n\{\psi_{S,t}k_h\}}{\mathbb{P}_n\{k_h\}} \equiv \mathbb{P}_{n,\cdot|k_h}\psi_{S,t}, \tag{S10}$$

where $P$ is the population average of function values, i.e., $Pf = \int f(h)dP(h)$, $\mathbb{P}_n$ is the sample average of function values, i.e., $\mathbb{P}_n f = \frac{1}{n}\sum_{i=1}^n f(H_i)$, $S_0 \in \Theta$ is the true survival probability of failure from any cause, $\delta_h(h') = 1(h' = h)$ is the unnormalized Dirac measure, and $k_h$ is the unnormalized tree kernel. Define $k_h(h') = 1(h' \in \mathcal{L}(h))$, where $\mathcal{L}(h)$ is the terminal node of the tree that contains the point $h$. The term 'unnormalized' refers to not being multiplied by the sample (or population) size. Let $\widehat{S}_n$ be the kernel-conditional Kaplan-Meier estimator, the Kaplan-Meier applied to the data with weights indicated by the tree kernel, $k_h$. Then, by Lemma 1 and Propositions 1 to 4 stated below, $\lim_{n\to\infty} \left\|\widehat{S}_n - S_0\right\|_\infty \to_p 0$. Since $\widehat{S}$ is an average of the trees, we have shown $\lim_{n\to\infty} \left\|\widehat{S} - S_0\right\|_\infty \to_p 0$.

**Proposition 1.** $\Psi(S_0) = 0$ *for* $S_0 \in \Theta$

**Proposition 2.** *Assume Assumptions 4-8. As* $n \to \infty$, $\left\|\Psi(\widehat{S}_n)\right\|_L \to 0$ *implies* $\left\|\widehat{S}_n - S_0\right\|_\infty \to 0$.

**Proposition 3.** *Assume Assumptions 4-8. As* $n \to \infty$, $\left\|\Psi_n(\widehat{S}_n)\right\|_L \to 0$ *in probability.*

**Proposition 4.** *Assume Assumptions 4-8. As* $n \to \infty$, $\sup_{S\in\Theta_n} \|\Psi_n(S) - \Psi(S)\|_L \to 0$ *in probability.*

*Proof of Proposition 1.* The numerator of $\Psi(S_0)$ is

$$\begin{aligned}
E(\delta_h \psi_{S_0,t}) &= E\left[\delta_h\left(I(X>t) + (1-\delta)I(X\le t)\frac{S_0(t\mid H)}{S_0(X\mid H)} - S_0(t\mid H)\right)\right] \\
&= E_{\delta_h} E_{\delta_h\psi_{S_0,t}|\delta_h}\left[\delta_h I(X>t) + \delta_h(1-\delta)I(X\le t)\frac{S_0(t\mid H)}{S_0(X\mid H)} - \delta_h S_0(t\mid H)\middle|\delta_h\right] \\
&= E_{\delta_h}\left(\delta_h I(X>t)\right) + E_{\delta_h}\left(\delta_h(1-\delta)I(X\le t)E_{\delta_h\psi_{S_0,t}|\delta_h}\left[\frac{S_0(t\mid H)}{S_0(X\mid H)}\middle|\delta_h\right]\right) \\
&\qquad - E_{\delta_h}\left(\delta_h E_{\delta_h\psi_{S_0,t}|\delta_h}\left[S_0(t\mid H)\middle|\delta_h\right]\right) \\
&= \Pr(\delta_h=1)\left\{E_{\delta_h\psi_{S_0,t}|\delta_h=1}\left[I(X>t)\right]\right. && (1.1) \\
&\qquad + E_{\delta_h\psi_{S_0,t}|\delta_h=1}\left[(1-\delta)I(X\le t)\frac{S_0(t\mid H)}{S_0(X\mid H)}\right] && (1.2) \\
&\qquad \left. - E_{\delta_h\psi_{S_0,t}|\delta_h=1}\left[S_0(t\mid H)\right]\right\} \\
&= \Pr(\delta_h=1)\left\{S_0(t|H=h)[1-G(t|H=h)]\right. \\
&\qquad \left. + S_0(t|H=h)G(t|H=h) - S_0(t|H=h)\right\} \\
&= \Pr(\delta_h=1)\left\{S_0(t|H=h) - S_0(t|H=h)G(t|H=h)\right. \\
&\qquad \left. + S_0(t|H=h)G(t|H=h) - S_0(t|H=h)\right\} = 0,
\end{aligned}$$

where $G(c) \equiv F_C(c)$ is the cumulative distribution function (CDF) for censoring time, and (1.1) and (1.2) are derived below.

For (1.1), $E(I(X > t_0)) = \Pr(X > t_0) = \Pr(T > t_0, C > t_0) = \Pr(T > t_0)\Pr(C >$

$t_0) = S_0(t_0)(1 - G(t_0))$. For (1.2):

$$
\begin{aligned}
&E_{\delta_h \psi_{S_0,t}|\delta_h=1}\left[(1-\delta)I(X \le t_0)I(S_0(X) > 0)\frac{S_0(t_0)}{S_0(X)}\right] \\
&= \int_0^{t_0}\int_c^{\infty} \frac{S_0(t_0)}{S_0(c)} f_{T,C}(t,c)\,dt\,dc \\
&= S_0(t_0)\int_0^{t_0} \frac{1}{S_0(c)} f_C(c) F_T(t)\Big|_c^{\infty}\,dc \\
&= S_0(t_0)\int_0^{t_0} \frac{1}{S_0(c)} f_C(c)\Big[F_T(\infty) - F_T(c)\Big]\,dc \\
&= S_0(t_0)\int_0^{t_0} f_C(c)\,dc \\
&= S_0(t_0)F_C(c)\Big|_0^{t_0} = S_0(t_0)[F_C(t_0) - F_C(0)] \equiv S_0(t_0)G(t_0).
\end{aligned}
$$

□

*Proof of Proposition 2.* By hypothesis, $\left\|\Psi(\widehat{S}_n)\right\|_L = \sup_{\substack{t\in(0,\tau] \\ h\in\mathcal{H}}} \left|P_{\cdot|h}\psi_{\widehat{S}_n,t}\right| \to 0$. We want to show that $P_{\cdot|h}\psi_{\widehat{S}_n,t} \to 0$ uniformly over $h \in \mathcal{H}$ and $t \in [0,\tau]$ implies $\sup_{t,h} u_n(t \mid h) \to 0$ for $u_n$ defined below. We know

$$
P_{\cdot|h}\psi_{\widehat{S}_n,t} = P_{\cdot|h}\left\{I(X > t) + (1-\delta_i)(X \le t)\frac{\widehat{S}_n(t)}{\widehat{S}_n(X)} - \widehat{S}_n(t)\right\}. \tag{S11}
$$

We individually examine each term in (S11). For the first term, because $T \perp C$, we have

$$
\begin{aligned}
P_{\cdot|h}\left(I(X > t)\right) &= E\left(I(X > t) \mid H = h\right) = \Pr(X > t \mid H = h) = \Pr(T > t, C > t \mid H = h) \\
&= \Pr(T > t \mid H = h)\Pr(C > t \mid H = h) = S_0(t \mid H = h)\left(1 - G(t \mid H = h)\right).
\end{aligned} \tag{S11.1}
$$

For the second term,

$$
P_{\cdot|h}\left\{(1-\delta)I(X \le t)\frac{\widehat{S}_n(t)}{\widehat{S}_n(X)}\right\} = E\left((1-\delta)I(X \le t)\frac{\widehat{S}_n(t)}{\widehat{S}_n(X)}\middle| H = h\right)
$$

$$= E\left( I(C \le t)\frac{\widehat{S}_n(t)}{\widehat{S}_n(C)} \middle| H = h, \delta = 0 \right)$$

$$= \int_0^t \int_c^\infty \frac{\widehat{S}_n(t \mid H = h)}{\widehat{S}_n(C \mid H = h)} f_{T,C}(u, c \mid H = h)\, du\, dc$$

$$= \int_0^t \frac{1}{\widehat{S}_n(c \mid H = h)} \int_c^\infty \widehat{S}_n(t \mid H = h) f_T(u \mid H = h)\, du\, f_C(c \mid H = h)\, dc$$

$$= \int_0^t \frac{f_C(c \mid H = h) S_0(c \mid H = h)}{\widehat{S}_n(c \mid H = h)}\, dc\, \widehat{S}_n(t \mid H = h), \tag{S11.2}$$

where the last equality holds because

$$\int_c^\infty \widehat{S}_n(t \mid H = h)\, f_T(u \mid H = h)\, du = \widehat{S}_n(t \mid H = h)\, F_T(u)\Big|_c^\infty = \widehat{S}_n(t \mid H = h)\,(1 - F_T(c))$$

$$= \widehat{S}_n(t \mid H = h)\,(1 - S_0(c)).$$

The last term is

$$P_{\cdot|h}\{-\widehat{S}_n(t)\} = -\widehat{S}_n(t \mid H = h). \tag{S11.3}$$

Combining (S11.1), (S11.2), (S11.3), we have

$$
\begin{aligned}
P_{\cdot|h}\psi_{\widehat{S}_n,t} &= S_0(t \mid H=h)\,(1 - G(t \mid H=h)) \\
&\qquad + \int_0^t \frac{f_C(c \mid H=h) S_0(c \mid H=h)}{\widehat{S}_n(c \mid H=h)}\, dc\, \widehat{S}_n(t \mid H=h) - \widehat{S}_n(t \mid H=h) \\
&= S_0(t \mid H=h) - \widehat{S}_n(t \mid H=h) - S_0(t \mid H=h) G(t \mid H=h) \\
&\qquad + \widehat{S}_n(t \mid H=h) \int_0^t \frac{f_C(c \mid H=h) S_0(c \mid H=h)}{\widehat{S}_n(c \mid H=h)}\, dc \\
&= \frac{\widehat{S}_n(t \mid H=h)}{\widehat{S}_n(t \mid H=h)} S_0(t \mid H=h) - \widehat{S}_n(t \mid H=h) \\
&\qquad - \frac{\widehat{S}_n(t \mid H=h)}{\widehat{S}_n(t \mid H=h)} S_0(t \mid H=h) G(t \mid H=h) \\
&\qquad + \widehat{S}_n(t \mid H=h) G(t \mid H=h) - \widehat{S}_n(t \mid H=h) G(t \mid H=h) \\
&\qquad + \widehat{S}_n(t \mid H=h) \int_0^t \frac{S_0(c \mid H=h)}{\widehat{S}_n(c \mid H=h)}\, dG(c \mid H=h) \\
&= \widehat{S}_n(t \mid H=h) \Bigg\{ \frac{S_0(t \mid H=h)}{\widehat{S}_n(t \mid H=h)} - 1 - \left[ \frac{S_0(t \mid H=h)}{\widehat{S}_n(t \mid H=h)} - 1 \right] G(t \mid H=h) \\
&\qquad\qquad - G(t \mid H=h) + \int_0^t \frac{S_0(c \mid H=h)}{\widehat{S}_n(c \mid H=h)}\, dG(c \mid H=h) \Bigg\}.
\end{aligned}
$$

Now, letting $\epsilon_n(t \mid H=h) = \frac{S_0(t|H=h)}{\widehat{S}_n(t|H=h)} - 1$, and noting that

$$
\begin{aligned}
&\int_0^t \frac{S_0(c \mid H=h)}{\widehat{S}_n(c \mid H=h)}\, dG(c \mid H=h) - G(t \mid H=h) \\
&\quad = \int_0^t \frac{S_0(c \mid H=h)}{\widehat{S}_n(c \mid H=h)}\, dG(c \mid H=h) - \int_0^t 1\, dG(c \mid H=h) \\
&\quad = \int_0^t \left[ \frac{S_0(c \mid H=h)}{\widehat{S}_n(c \mid H=h)} - 1 \right] dG(c \mid H=h) = \int_0^t \epsilon_n(c \mid H=h)\, dG(c \mid H=h),
\end{aligned}
$$

we have $P_{\cdot|h}\psi_{\widehat{S}_n,t}$

$$= \widehat{S}_n(t \mid H=h)\Bigg\{\epsilon_n(t \mid H=h)(1-G(t \mid H=h)) + \int_0^t \frac{S_0(c \mid H=h)}{\widehat{S}_n(c \mid H=h)}\, dG(c \mid H=h) - G(t \mid H=h)\Bigg\}$$

$$= \widehat{S}_n(t \mid H=h)\Bigg\{\epsilon_n(t \mid H=h)(1-G(t \mid H=h)) + \int_0^t \epsilon_n(c \mid H=h)\, dG(c \mid H=h)\Bigg\}.$$

Setting $u_n \equiv \epsilon_n(t \mid H=h)(1-G(t \mid H=h)) + \int_0^t \epsilon_n(c \mid H=h)\, dG(c \mid H=h)$, and taking the derivative on both sides, we get

$$\begin{aligned} du_n(t \mid H=h) &= d\Bigg[\epsilon_n(t \mid H=h)(1-G(t \mid H=h)) + \int_0^t \epsilon_n(c \mid H=h)\, dG(c \mid H=h)\Bigg] \\ &= d\epsilon_n(t)(1-G(t \mid H=h)) + \Big(d(1-G(t \mid H=h))\Big)\epsilon_n(t \mid H=h) \\ &\quad + \epsilon_n(t \mid H=h) dG(t \mid H=h) \\ &= d\epsilon_n(t)(1-G(t \mid H=h)) - dG(t \mid H=h)\, \epsilon_n(t \mid H=h) \\ &\quad + \epsilon_n(t \mid H=h) dG(t \mid H=h) \\ &= d\epsilon_n(t)(1-G(t \mid H=h)) \end{aligned}$$

$$\implies d\epsilon_n(t \mid H=h) = \frac{du_n(t)}{1-G(t \mid H=h)}.$$

Using integration by parts, we have

$$\begin{aligned} \epsilon_n(t \mid H=h) &= \int_0^t \frac{du_n(t)}{1-G(t \mid H=h)} \\ &= \frac{u_n(u \mid H=h)}{1-G(u \mid H=h)}\Bigg|_0^t - \int_0^t \frac{u_n(u \mid H=h)\, dG(u \mid H=h)}{(1-G(u \mid H=h))^2} \\ &= \frac{u_n(t \mid H=h)}{1-G(t \mid H=h)} - \int_0^t \frac{u_n(u \mid H=h)\, dG(u \mid H=h)}{(1-G(u \mid H=h))^2}. \end{aligned}$$

By Assumption 4 , $1-G(t \mid H=h) \geq 1-G(\tau \mid H=h) > 0$. Thus, $u_n(t \mid H=h) \to 0$ implies $\epsilon_n(t \mid H=h) \to 0$. Since $u_n(t \mid H=h) \to 0$ by hypothesis, we have $\frac{S_0(t|H=h)}{\widehat{S}_n(t|H=h)} - 1 \to 0$. Thus, $\left\|\widehat{S}_n - S_0\right\|_\infty \to 0$.

□

*Proof of Proposition 3.* We utilize proof by induction on $t \in (0,\tau]$. $\sum_{i=1}^n \psi_{\widehat{S}_n,t=0}(X_i) = \sum_{i=1}^n \Big\{I(X > 0) + (1-\delta_i)I(X \leq 0)\frac{S(0|H)}{S(X|H)} - S(0 \mid H)\Big\} = \sum_{i=1}^n \Big\{I(X > 0) - S(0 \mid$

$H)\Big\} = \sum_{i=1}^n \Big\{1 - 1\Big\} = 0$. Then for all $0 < t \le \tau$, if $\sum_{i=1}^n \psi_{\widehat{S}_n,t^-}(X_i) = 0$ implies $\sum_{i=1}^n \psi_{S,t}(X_i) = 0$, the desired result holds.

We re-write $\psi_{\widehat{S}_n,t}$ in terms of counting notation, using jumps such that $I(X_i \le t) = I(X_i \le t^-) + dI(X_i \le t)$ and $I(X_i > t) = I(X_i > t^-) + dI(X_i > t)$.

$$\begin{aligned}
\sum_{i=1}^n \psi_{\widehat{S}_n,t} &= \sum_{i=1}^n I(X_i > t) + \sum_{i=1}^n \left[(1-\delta_i) I(X_i \le t) \frac{\widehat{S}_n(t \mid H)}{\widehat{S}_n(X_i \mid H)} - \widehat{S}_n(t \mid H)\right] \\
&= \sum_{i=1}^n I(X_i > t) + \sum_{i=1}^n \left[\frac{(1-\delta_i) I(X_i \le t)}{\widehat{S}_n(X_i \mid H)} - 1\right] \widehat{S}_n(t \mid H) \\
&= \sum_{i=1}^n I(X_i > t^-) + \sum_{i=1}^n dI(X_i > t) \\
&\quad + \sum_{i=1}^n \left[\frac{(1-\delta_i) I(X_i \le t^-)}{\widehat{S}_n(X_i \mid H)} - 1\right] \widehat{S}_n(t \mid H) \\
&\quad + \sum_{i=1}^n \left[\frac{(1-\delta_i) dI(X_i \le t)}{\widehat{S}_n(X_i \mid H)}\right] \widehat{S}_n(t \mid H) \\
&= \sum_{i=1}^n \left[I(X_i > t^-) + dI(X_i > t)\right] \\
&\quad + \sum_{i=1}^n \left[\frac{(1-\delta_i) I(X_i \le t^-)}{\widehat{S}_n(X_i \mid H)} - 1\right] \widehat{S}_n(t^- \mid H) \\
&\quad + \sum_{i=1}^n \left[\frac{(1-\delta_i) I(X_i \le t^-)}{\widehat{S}_n(X_i \mid H)} - 1\right] \widehat{S}_n(t^- \mid H) \frac{\sum_{j=1}^n \delta_j dI(X_j > t)}{\sum_{j=1}^n I(X_j > t^-)} \\
&\quad + \sum_{i=1}^n \left[\frac{(1-\delta_i) I(X_i \le t)}{\widehat{S}_n(X_i \mid H)}\right] \widehat{S}_n(t \mid H),
\end{aligned}$$

where the second equality is because

$$\sum_{i=1}^{n}\left[\frac{(1-\delta_i)I(X_i \le t^-)}{\widehat{S}_n(X_i \mid H)} - 1\right]\widehat{S}_n(t \mid H)$$

$$= \sum_{i=1}^{n}\left(\frac{(1-\delta_i)[I(X_i \le t^-) + dI(X_i \le t)]}{\widehat{S}_n(X_i \mid H)} - 1\right)\widehat{S}_n(t \mid H)$$

$$= \sum_{i=1}^{n}\left(\frac{(1-\delta_i)I(X_i \le t^-)}{\widehat{S}_n(X_i \mid H)} + \frac{(1-\delta_i)dI(X_i \le t)}{\widehat{S}_n(X_i \mid H)} - 1\right)\widehat{S}_n(t \mid H)$$

$$= \sum_{i=1}^{n}\left(\frac{(1-\delta_i)I(X_i \le t^-)}{\widehat{S}_n(X_i \mid H)} - 1\right)\widehat{S}_n(t \mid H) + \left(\frac{(1-\delta_i)dI(X_i \le t)}{\widehat{S}_n(X_i \mid H)}\right)\widehat{S}_n(t \mid H),$$

and the third equality is because $\widehat{S}_n(t \mid H) = \widehat{S}_n(t^- \mid H) + \widehat{S}_n(t^- \mid H)\frac{\sum_{i=1}^{n}\delta_i dI(X_i > t)}{\sum_{i=1}^{n} I(X_i > t^-)}$ from the definition of modified Kaplan-Meier estimator in Cho et al. (2023).

Since we know by hypothesis that $\sum_{i=1}^{n}\psi_{\widehat{S}_n,t^-} = \sum_{i=1}^{n}[I(X_i > t^-) + (1-\delta_i)I(X_i \le t^-)\frac{\widehat{S}_n(t^- \mid H)}{\widehat{S}_n(X_i \mid H)} - \widehat{S}_n(t \mid H)] = 0$, we re-write $\psi_{\widehat{S}_n,t}$ into smaller components.

$$\sum_{i=1}^{n}\psi_{\widehat{S}_n,t} = \sum_{i=1}^{n} dI(X_i > t)$$

$$+\sum_{i=1}^{n} I(X_i > t^-) + \sum_{i=1}^{n}\left[\frac{(1-\delta_i)I(X_i \le t^-)}{\widehat{S}_n(X_i \mid H)} - 1\right]\widehat{S}_n(t^- \mid H) \tag{3.1}$$

$$+\sum_{i=1}^{n}\left[\frac{(1-\delta_i)I(X_i \le t^-)}{\widehat{S}_n(X_i \mid H)} - 1\right]\widehat{S}_n(t^- \mid H)\frac{\sum_{j=1}^{n}\delta_j dI(X_j > t)}{\sum_{j=1}^{n} I(X_j > t^-)} \tag{3.2}$$

$$+\sum_{i=1}^{n}\left[\frac{(1-\delta_i)I(X_i \le t)}{\widehat{S}_n(X_i \mid H)}\right]\widehat{S}_n(t \mid H) \tag{3.3}$$

$$= \sum_{i=1}^{n} dI(X_i > t) - \sum_{i=1}^{n}\delta_i dI(X_i > t) + \sum_{i=1}^{n}(1-\delta_i)I(X_i = t)$$

$$= -\sum_{i=1}^{n} I(X_i = t) + \sum_{i=1}^{n}\delta_i I(X_i = t) + \sum_{i=1}^{n}(1-\delta_i)I(X_i = t)$$

$$= 0,$$

where (3.1) $= \sum_{i=1}^{n}\psi_{\widehat{S}_n,t^-} = 0$ by hypothesis, and the derivations from (3.2) and (3.3) to the next line are given below.

We use the hypothesis for (3.2). Since $\sum_{i=1}^{n}\psi_{\widehat{S}_n,t^-} = 0$, then $\sum_{i=1}^{n}\Bigg[\frac{(1-\delta_i)I(X_i \le t^-)}{\widehat{S}_n(X_i \mid H)} -$ $1\Big]\widehat{S}_n(t^- \mid H) = -\sum_{i=1}^n I(X_i > t^-)$, so $\sum_{i=1}^n \left[\frac{(1-\delta_i)I(X_i \le t^-)}{\widehat{S}_n(X_i|H)} - 1\right]\widehat{S}_n(t^- \mid H)\frac{\sum_{j=1}^n \delta_j dI(X_j>t)}{\sum_{j=1}^n I(X_j>t^-)} = -\sum_{i=1}^n I(X_i > t^-)\frac{\sum_{j=1}^n \delta_j dI(X_j>t)}{\sum_{j=1}^n I(X_j>t^-)} = -\sum_{i=1}^n \delta_i dI(X_i > t)$.

For (3.3), this term is not equal to 0 when $\delta_i = 0$ and $dI(X_i \le t) = 1$. We note that $dI(X_i \le t) = I(X_i = t) = 1$ when $t = X_i$, and 0 otherwise. In other words, when $dI(X_i \le t) = 1$, $\widehat{S}_n(X_i) = \widehat{S}_n(t)$. Because $I(X_i \le t)$ is increasing and the jump at $t = X_i$ is positive (from 0 to 1), the derivative is positive. Then, when (3.3) is not 0, $\sum_{i=1}^n \left[\frac{(1-\delta_i)I(X_i \le t)}{\widehat{S}_n(X_i|H)}\right]\widehat{S}_n(t \mid H) = \sum_{i=1}^n \left[\frac{(1-\delta_i)I(X_i = t)}{\widehat{S}_n(t|H)}\right]\widehat{S}_n(t \mid H) = \sum_{i=1}^n (1-\delta_i)I(X_i = t)$. We note that, using the same logic, $dI(X_i > t) = -I(X_i = t)$, since the function $I(X_i > t)$ is decreasing and the jump at $t = X_i$ goes from 1 to 0. □

Proposition 5 is needed to prove Proposition 4 below. This proposition shows that tree kernels (and, consequently, the forest kernels) are Donsker, and that the $\Psi(S)$ function is also Donsker. The proof can be found in Cho et al. (2023). We use entropy calculation, the Donsker theorem, empirical process theory, and Assumptions 4-8 to establish uniform consistency in Proposition 4.

**Proposition 5.** *(Bounded entropy integral of the tree and forest kernels). The collections of the unnormalized tree and forest kernel functions are Donsker, where the tree kernels $k_{tree}(\cdot)$ are axis-aligned random hyper-rectangles, and the forest kernels $k_{forest}(\cdot)$ are the mean of arbitrarily many ($n_{tree}$) tree kernels.*

*Proof of Proposition 4.* We first show that the class of functions $\{\psi_{S,t} : S \in \Theta_n, t \in [0,\tau]\}$ is Donsker. Since the components of $\psi_{S,t}$ are uniformly bounded Donsker classes, by Lemma 9.10 of Kosorok (2008), the class of functions is Donsker.

With Proposition 5 and the class of products of bounded Donsker classes being Donsker, all of $\{\psi_{S,t}\delta_h : S \in \Theta_n, t \in [0,\tau], h \in \mathcal{H}\}$ and $\{\psi_{S,t}k_h : S \in \Theta_n, t \in [0,\tau], h \in \mathcal{H}\}$ are Donsker. By Assumption 5, $\sup_h \mathbb{P}_n k_h \bowtie n^{\beta-1}$. Then,

$$
\begin{aligned}
\sup_{\mathrm{S}\in\Theta_\mathrm{n}} \|\Psi_n(S) - \Psi(S)\|_{\mathbb{L}} &= \sup_{\mathrm{S}\in\Theta_\mathrm{n}} \|\mathbb{P}_n\psi_{S,t}k_h - P\psi_{S,t}\delta_h\|_{\mathbb{L}} \\
&= \sup_{\mathrm{S}\in\Theta_\mathrm{n}} \left\|\mathbb{P}_{n,\cdot|k_h}\psi_{S,t}\psi_{S,t} - P_{\cdot|k_h}\psi_{S,t} + P_{\cdot|k_h}\psi_{S,t} - P_{\cdot|h}\psi_{S,t}\right\|_{\mathbb{L}} \\
&\le \sup_{\substack{S\in\Theta_n \\ t\in(0,\tau] \\ h\in\mathcal{H}}} \left|\frac{\mathbb{P}_n\{\psi_{S,t}k_h\}}{\mathbb{P}_n\{k_h\}} - \frac{P\{\psi_{S,t}k_h\}}{P\{k_h\}}\right| && \text{(S12)} \\
&+ \sup_{\substack{S\in\Theta_n \\ t\in(0,\tau] \\ h\in\mathcal{H}}} \left|\frac{P\{\psi_{S,t}k_h\}}{P\{k_h\}} - \frac{P\{\psi_{S,t}\delta_h\}}{P\{\delta_h\}}\right| && \text{(S13)} \\
&= o_P(1).
\end{aligned}
$$

To see this, we show each component (S12) and (S13) are $o_P(1)$.

We can break (S12) down further.

$$(S12) = \sup_{\substack{S\in\Theta_n \\ t\in(0,\tau] \\ h\in\mathcal{H}}} \left| \frac{\mathbb{P}_n\{\psi_{S,t}k_h\} - P\{\psi_{S,t}k_h\}}{\mathbb{P}_n\{k_h\}} \right| \tag{S12.1}$$

$$+ \sup_{\substack{S\in\Theta_n \\ t\in(0,\tau] \\ h\in\mathcal{H}}} \left| P\{\psi_{S,t}k_h\} \left[ \frac{1}{\mathbb{P}_n k_h} - \frac{1}{P k_h} \right] \right|. \tag{S12.2}$$

Because $(S12.1) \leq \frac{O_p(n^{-1/2})}{n^{\beta-1}}$ and $(S12.2) = \frac{P\{\psi_{S,t}k_h\}}{Pk_h}\frac{(\mathbb{P}_n-P)k_h}{\mathbb{P}_n k_h} \leq \frac{O_P(n^{-1/2})}{n^{\beta-1}}$, we obtain $S12 = o_P(1)$.

We now show (S13) $= o_P(1)$. We denote $S(t \mid k_h)$ as shorthand for $S(t \mid H \in \mathbb{L}(h))$ and note that $S(t \mid k_h) = S(t \mid h)$ for all $S \in \Theta_n$. Then for all $S \in \Theta_n$, $t \in [0, \tau]$ and $h \in \mathcal{H}$,

$$
\begin{aligned}
&\left|P_{\cdot|k_h}\psi_{S,t} - P_{\cdot|h}\psi_{S,t}\right| \\
&= \Bigg| S_0(t \mid k_h)(1 - G(t \mid k_h)) - S(t \mid k_h) + S(t \mid k_h)\int_0^t \frac{S_0(u \mid k_h)}{S(u \mid k_h)} dG(u \mid k_h) \\
&\qquad - S_0(t \mid h)(1 - G(t \mid h)) + S(t \mid h) - S(t \mid h)\int_0^t \frac{S_0(u \mid h)}{S(u \mid h)} dG(u \mid h) \Bigg| \\
&= \Bigg| S_0(t \mid k_h)(1 - G(t \mid k_h)) - S_0(t \mid h)(1 - G(t \mid h)) \\
&\qquad + S(t \mid h)\int_0^t \left[\frac{S_0(u \mid k_h)}{S(u \mid h)} dG(u \mid k_h) - \frac{S_0(u \mid h)}{S(u \mid h)} dG(u \mid h)\right] \Bigg| \\
&\leq |S_0(t \mid k_h)(1 - G(t \mid k_h)) - S_0(t \mid h)(1 - G(t \mid h))| \\
&\qquad + \Bigg| S(t \mid h)\int_0^t \left[\frac{S_0(u \mid k_h)}{S(u \mid h)} dG(u \mid k_h) - \frac{S_0(u \mid h)}{S(u \mid h)} dG(u \mid h) \right. \\
&\qquad \left. + \frac{S_0(u \mid k_h)}{S(u \mid h)} dG(u \mid h) - \frac{S_0(u \mid k_h)}{S(u \mid h)} dG(u \mid h)\right] \Bigg| \\
&\leq |S_0(t \mid k_h)(1 - G(t \mid k_h)) - S_0(t \mid h)(1 - G(t \mid h))| \\
&\qquad + \Bigg| S(t \mid h)\int_0^t \left\{\frac{S_0(u \mid k_h)}{S(u \mid h)}\Big[dG(u \mid k_h) - dG(u \mid h)\Big] \right. \\
&\qquad \left. + \left[\frac{S_0(u \mid k_h) - S_0(u \mid h)}{S(u \mid h)}\right] dG(u \mid h)\right\} \Bigg| \\
&\leq \Big| S_0(t \mid k_h)(1 - G(t \mid k_h)) - S_0(t \mid h)(1 - G(t \mid h)) \Big| && \text{(S13.1)} \\
&\qquad + \left| S(t \mid h)\int_0^t \frac{S_0(u \mid k_h)}{S(u \mid h)}\Big[dG(u \mid k_h) - dG(u \mid h)\Big] \right| && \text{(S13.2)} \\
&\qquad + \left| S(t \mid h)\int_0^t \left[\frac{S_0(u \mid k_h) - S_0(u \mid h)}{S(u \mid h)}\right] dG(u \mid h) \right|. && \text{(S13.3)}
\end{aligned}
$$

$(S13.1)$, $(S13.2)$, and $(S13.3)$ can be further bounded by using Assumption 7.

$$
\begin{aligned}
(S13.1) &= \Big| S_0(t \mid k_h)(1 - G(t \mid k_h)) - S_0(t \mid h)(1 - G(t \mid h)) \Big| \\
&= \Big| S_0(t \mid k_h)(1 - G(t \mid k_h)) - S_0(t \mid h)(1 - G(t \mid h)) \\
&\qquad + S_0(t \mid k_h)(1 - G(t \mid h)) - S_0(t \mid k_h)(1 - G(t \mid h)) \Big| \\
&= \Big| S_0(t \mid k_h)\{(1 - G(t \mid k_h)) - (1 - G(t \mid h))\} + (1 - G(t \mid h))\{S_0(t \mid k_h) - S_0(t \mid h)\} \Big| \\
&= S_0(t \mid k_h)\Big| (G(t \mid h) - G(t \mid k_h)) \Big| + (1 - G(t \mid h))\Big| S_0(t \mid k_h) - S_0(t \mid h) \Big| \\
&\le L_G \sup_{\mathrm{h}' \in \mathcal{H}:\mathrm{k_h}(\mathrm{h}')>0} \|h - h'\|_\infty + L_S \sup_{\mathrm{h}' \in \mathcal{H}:\mathrm{k_h}(\mathrm{h}')>0} \|h - h'\|_\infty ,
\end{aligned}
$$

where the last inequality holds because $S_0(t \mid k_h) \le 1$ and $1 - G(t \mid h) \le 1$.

To bound (S13.2) further, we use $dG(u) \equiv dF_C(u) = f_C(u)du$.

$$
\begin{aligned}
(S13.2) &= \left| S(t \mid h) \int_0^t \frac{S_0(u \mid k_h)}{S(u \mid h)} \Big[ dG(u \mid k_h) - dG(u \mid h) \Big] \right| \\
&\le \int_0^t S_0(u \mid k_h) \frac{S(t \mid h)}{S(u \mid h)} \Big| dG(u \mid k_h) - dG(u \mid h) \Big| \\
&\le \int_0^t \Big| dG(u \mid k_h) - dG(u \mid h) \Big| = \int_0^t \Big| f_C(u \mid k_h) - f_C(u \mid h) \Big| du \\
&\le \int_0^t 2L_G \sup_{\mathrm{h}' \in \mathcal{H}:\mathrm{k_h}(\mathrm{h}')>0} \|h - h'\|_\infty \, du \\
&\le 2t \, L_G \sup_{\mathrm{h}' \in \mathcal{H}:\mathrm{k_h}(\mathrm{h}')>0} \|h - h'\|_\infty ,
\end{aligned}
$$

where the second inequality holds by bringing in the absolute value inside the integral, the third inequality holds because $S_0(u \mid k_h) \le 1$ and $u \le t \implies S(u) \le S(t) \implies \frac{S(t|h)}{S(u|h)} \le 1$, and the fourth inequality holds by derivative of Lipschitz function.

We now further bound (S13.3).

$$
\begin{aligned}
(S13.3) &= \left| \int_0^t \frac{S(t \mid h)}{S(u \mid h)} \Big[ S_0(u \mid k_h) - S_0(u \mid h) \Big] dG(u \mid h) \right| \\
&\le \sup_t \sup_h \Big| S_0(t \mid k_h) - S_0(t \mid h) \Big| \int_0^t dG(u \mid h) \\
&\le L_S \sup_{\mathrm{h}' \in \mathcal{H}:\mathrm{k_h}(\mathrm{h}')>0} \|h - h'\|_\infty ,
\end{aligned}
$$

where the first inequality is obtained because $\frac{S(t|h)}{S(u|h)} \le 1$ and the double supremum comes from maximizing $S_0 - S$ to obtain a constant that can be take outside of the integral, and the second inequality is due to $\int_0^t dG(u \mid h) = G(t \mid h) - G(0 \mid h) = G(t \mid h) \le 1$. The last line drops $\sup_t$ because $\|h - h'\|_\infty$ is no longer a function of $t$.

As a result, we have

$$(S13) = \sup_{\substack{S\in\Theta_n \\ t\in(0,\tau] \\ h\in\mathcal{H}}} \left|P_{\cdot|k_h}\psi_{S,t} - P_{\cdot|h}\psi_{S,t}\right| \le \{(2\tau+1)L_G + 2L_S\} \sup_{\mathrm{h}'\in\mathcal{H}:\mathrm{k_h(h')}>0} \|h - h'\|_\infty .$$

Following similar arguments as in Meinshausen (2006) and Cho et al. (2023), with Assumptions 5, 6, 8, it can be shown that $\sup_{\mathrm{h}'\in\mathcal{H}:\mathrm{k_h(h')}>0} \|h - h'\|_\infty = o_P(1)$.

□

□

# E Proof of Theorem 3

*Proof of Theorem 3.* Our estimator is

$$\widehat{F}_n(t \mid h) = \int_0^t \widehat{S}_n^h(u^-)\frac{\mathbb{P}_n k_h(H) dN_j^h(u)}{\mathbb{P}_n k_h(H) Y(u)}.$$

For simplicity, we drop the $h$ superscript for any survival function when it does not cause confusion (i.e., $\widehat{S}_n^h \equiv \widehat{S}_n$), but keep in mind that the survival functions in this proof use both arguments $t$ and $h$, since the RCIF is built based on $\{H_i, \delta_i, \delta_{ji}\}_{i=1}^n$. We do the same for $N_j(t)$ such that $N_j^h(t) \equiv N_j(t)$ where $N_j^h(t)$ are the events due to cause-$j$ at time $t$ with history $h$.

$$\begin{aligned}
\Longrightarrow \widehat{F}_n(t \mid h) &\equiv \int_0^t \widehat{S}_n(u^-)\frac{\mathbb{P}_n k_h(H) dN_j(u)}{\mathbb{P}_n k_h(H) Y(u)} \\
&= \int_0^t \widehat{S}_n(u^-)\frac{\mathbb{P}_n k_h(H) dN_j(u)}{\mathbb{P}_n k_h(H) Y(u)} - \int_0^t S_0(u^-)\frac{\mathbb{P}_n k_h(H) dN_j(u)}{\mathbb{P}_n k_h(H) Y(u)} \\
&\quad + \int_0^t S_0(u^-)\frac{\mathbb{P}_n k_h(H) dN_j(u)}{\mathbb{P}_n k_h(H) Y(u)} \\
&= \int_0^t \left\{\widehat{S}_n(u^-) - S_0(u^-)\right\}\frac{\mathbb{P}_n k_h(H) dN_j(u)}{\mathbb{P}_n k_h(H) Y(u)} + \int_0^t S_0(u^-)\frac{\mathbb{P}_n k_h(H) dN_j(u)}{\mathbb{P}_n k_h(H) Y(u)}. \qquad \text{(S14)}
\end{aligned}$$

Let $d\widehat{\Lambda}_n^h(t) = \frac{\mathbb{P}_n k_h(H) dN_j(t)}{\mathbb{P}_n k_h(H) Y(t)}$ and $d\widetilde{\Lambda}_n^h(t) = \frac{P k_h(H) dN_j(t)}{P k_h(H) Y(t)}$. First, we obtain a bound for $\left|\int_0^t \left\{\widehat{S}_n(u^-) - S_0(u^-)\right\} d\widehat{\Lambda}_n^h(t)\right|$ for $t \in (0,\tau]$ and $h \in \mathcal{H}$:

$$\left|\int_0^t \left\{\widehat{S}_n(u^-) - \int_0^t S_0(u^-)\right\} d\widehat{\Lambda}_n^h(u)\right| \le \left\{\sup_{\substack{t\in(0,\tau] \\ h\in\mathcal{H}}} \left|\widehat{S}_n(t) - S_0(t)\right|\right\} \int_0^t d\widehat{\Lambda}_n^h(u),$$

where $\sup_{t\in(0,\tau]}\left|\widehat{S}_n(t)-S_0(t)\right|$ (with $t^-$ incorporated into $\sup_{t\in(0,\tau]}$) has the rate for Kaplan-Meier for the single-stage generalized random survival forest, and $\int_0^t d\widehat{\Lambda}_n^h(s) \le \widehat{\Lambda}_n^h(\tau)$ is uniformly bounded over $h$. Then, the first term of (S14) converges to 0.

We next examine the second term of (S14). We can re-write it by dividing both the numerator and denominator by $\mathbb{P}_n k_h(H)$ to get $\int_0^t S_0(u^-)\frac{\mathbb{P}_n k_h(H) dN_j(u)/\mathbb{P}_n k_h(H)}{\mathbb{P}_n k_h(H) Y(u)/\mathbb{P}_n k_h(H)}$. We will use Lemma 12.3 from Kosorok (2008). Since $\{\delta_j I(X\le t), t\ge 0\}$ and $\{I(X\ge t), t\ge 0\}$ are indicators and $dN_j$ is at most 1, they are bounded Donsker. Since $N_j$ and $Y$ are bounded Donsker, $k_h$ is bounded Donsker, and the product of bounded Donsker is Donsker, then

$$\sqrt{n}\,\Delta_n(h,t) \equiv \sqrt{n}\begin{pmatrix}\mathbb{P}_n k_h(H)N_j(t) - Pk_h(H)N_j(t)\\ \mathbb{P}_n k_h(H)Y(t) - Pk_h(H)Y(t)\\ \mathbb{P}_n k_h(H) - Pk_h(H)\end{pmatrix} \rightsquigarrow \begin{pmatrix}\mathbb{G}_1(h,t)\\ \mathbb{G}_2(h,t)\\ \mathbb{G}_3(h,t)\end{pmatrix}, \tag{S15}$$

where $\mathbb{G}_1$, $\mathbb{G}_2$, and $\mathbb{G}_3$ are tight Gaussian processes. Thus, $\sqrt{n}\,\Delta_n(h,t) = O_p^{(h,t)}(1)$, where $O_p^{(h,t)}(1)$ is a quantity $R_n(h,t)$ such that $\sup_{\substack{t\in(0,\tau]\\ h\in\mathcal{H}}}|R_n(h,t)| = O_p(1)$, which implies $\Delta_n(h,t) = O_p^{(h,t)}(n^{-1/2})$.

Since $S_0(t^-)$ is bounded away from 0, we have from (S15),

$$\sqrt{n}\begin{pmatrix}\mathbb{P}_n k_h(H)N_j(t) - Pk_h(H)N_j(t)\\ \frac{1}{S_0(t^-)}\mathbb{P}_n k_h(H)Y(t) - \frac{1}{S_0(t^-)}Pk_h(H)Y(t)\\ \mathbb{P}_n k_h(H) - Pk_h(H)\end{pmatrix} \rightsquigarrow \begin{pmatrix}\mathbb{G}_1(h,t)\\ \frac{1}{S_0(t^-)}\mathbb{G}_2(h,t)\\ \mathbb{G}_3(h,t)\end{pmatrix}. \tag{S16}$$

First, we examine $N_j$. From (S15), we have

$$\frac{\mathbb{P}_n k_h(H)N_j(t)}{\mathbb{P}_n k_h(H)} = \frac{Pk_h(H)N_j(t) + O_p^{(h,t)}(n^{-1/2})}{Pk_h(H) + O_p^{(h,t)}(n^{-1/2})}. \tag{S17}$$

Assumption 5 implies that $\inf_{h\in\mathcal{H}} Pk_h(H) \ge k_0 n^{\beta-1}$ for all $h\in\mathcal{H}$ where $k_0>0$ is some constant. Note that $n^{1-\beta}\times n^{-1/2} = n^{\frac{1}{2}-\beta}$. Multiplying $n^{1-\beta}$ to the numerator and denominator of the right-hand side of (S17), we obtain

$$\frac{\mathbb{P}_n k_h(H)N_j(t)}{\mathbb{P}_n k_h(H)} = \frac{n^{1-\beta}Pk_h(H)N_j(t) + O_p^{(h,t)}(n^{\frac{1}{2}-\beta})}{n^{1-\beta}Pk_h(H) + O_p^{(h,t)}(n^{\frac{1}{2}-\beta})}. \tag{S18}$$

The denominator of the right-hand side of (S18) is greater than or equal to $k_0 + o_p^{(h,t)}(1)$ because $n^{1-\beta}Pk_h(H) \ge n^{1-\beta}k_0 n^{\beta-1} = k_0$ and the residual error is going to zero since by Assumption 5, $\beta > \frac{1}{2}$. Therefore, we have

$$\frac{\mathbb{P}_n k_h(H)N_j(t)}{\mathbb{P}_n k_h(H)} = \frac{n^{1-\beta}Pk_h(H)N_j(t) + o_p^{(h,t)}(1)}{n^{1-\beta}Pk_h(H) + o_p^{(h,t)}(1)}. \tag{S19}$$

We note that

$$\frac{n^{1-\beta}Pk_h(H) + o_p^{(h,t)}(1)}{n^{1-\beta}Pk_h(H)} = 1 + \frac{o_p^{(h,t)}(1)}{n^{1-\beta}Pk_h(H)} = 1 + o_p^{(h,t)}(1), \tag{S20}$$

where the last equality in (S20) holds because $n^{1-\beta}Pk_h(H) \geq k_0 > 0$. Since $(1+o_p^{(h,t)}(1))^{-1} = 1 + o_p^{(h,t)}(1)$, we can take the inverse of (S20).

$$\Longrightarrow \frac{n^{1-\beta}Pk_h(H)}{n^{1-\beta}Pk_h(H) + o_p^{(h,t)}(1)} = 1 + o_p^{(h,t)}(1)$$

$$\Longrightarrow \frac{1}{n^{1-\beta}Pk_h(H) + o_p^{(h,t)}(1)} = \frac{1 + o_p^{(h,t)}(1)}{n^{1-\beta}Pk_h(H)}. \tag{S21}$$

To continue from (S19), multiply both sides of (S21) by $n^{1-\beta}Pk_h(H)N_j(t) + o_p^{(h,t)}(1)$.

$$\Longrightarrow \frac{n^{1-\beta}Pk_h(H)N_j(t) + o_p^{(h,t)}(1)}{n^{1-\beta}Pk_h(H) + o_p^{(h,t)}(1)} = \frac{n^{1-\beta}Pk_h(H)N_j(t) + o_p^{(h,t)}(1)}{n^{1-\beta}Pk_h(H)}\Big(1 + o_p^{(h,t)}(1)\Big).$$

We re-write the numerator as $n^{1-\beta}Pk_h(H)N_j(t) = n^{\frac{1}{2}-\beta}n^{\frac{1}{2}}Pk_h(H)N_j(t)$. Since $\sqrt{n}Pk_h(H)N_j(t) = O_p^{(h,t)}(1)$ and $n^{\frac{1}{2}-\beta} \to 0$, $n^{1-\beta}Pk_h(H)N_j(t)$ is bounded. The denominator is bounded away from 0 because $n^{1-\beta}Pk_h(H) \geq k_0 > 0$. This implies,

$$\begin{aligned}
\frac{\mathbb{P}_n k_h(H)N_j(t)}{\mathbb{P}_n k_h(H)} &= \frac{n^{1-\beta}Pk_h(H)N_j(t)}{n^{1-\beta}Pk_h(H)}\Big(1 + o_p^{(h,t)}(1)\Big) + o_p^{(h,t)}(1) \\
&= \frac{Pk_h(H)N_j(t)}{Pk_h(H)}\Big(1 + o_p^{(h,t)}(1)\Big) + o_p^{(h,t)}(1) \\
&= \frac{Pk_h(H)N_j(t)}{Pk_h(H)} + o_p^{(h,t)}(1) \quad \text{(because } \frac{Pk_h(H)N_j(t)}{Pk_h(H)} \text{ is bounded).}
\end{aligned}$$

Hence,

$$\frac{\mathbb{P}_n k_h(H)N_j(t)}{\mathbb{P}_n k_h(H)} = \frac{Pk_h(H)N_j(t)}{Pk_h(H)} + o_p^{(h,t)}(1).$$

Using similar arguments, we can show that the kernel version of $Y$, $\frac{1}{S_0(t^-)}\mathbb{P}_n k_h(H)Y(t)$ is consistent as well. Specifically, since $S_0(t)$ is bounded away from 0, similar to (S18), we have

$$\frac{\frac{1}{S_0(t^-)}\mathbb{P}_n k_h(H)Y(t)}{\mathbb{P}_n k_h(H)} = \frac{n^{1-\beta}\frac{1}{S_0(t^-)}Pk_h(H)Y(t) + O_p^{(h,t)}(n^{\frac{1}{2}-\beta})}{n^{1-\beta}Pk_h(H) + O_p^{(h,t)}(n^{\frac{1}{2}-\beta})} = \frac{n^{1-\beta}\frac{1}{S_0(t^-)}Pk_h(H)Y(t) + o_p^{(h,t)}(1)}{n^{1-\beta}Pk_h(H) + o_p^{(h,t)}(1)}.$$

Multiplying $n^{1-\beta}\frac{1}{S_0(t^-)}Pk_h(H)Y(t) + o_p^{(h,t)}(1)$ on the both sides of (S21) and simplifying the expressions, we have

$$\Longrightarrow \frac{n^{1-\beta}\frac{1}{S_0(t^-)}Pk_h(H)Y(t) + o_p^{(h,t)}(1)}{n^{1-\beta}Pk_h(H) + o_p^{(h,t)}(1)} = \frac{\frac{1}{S_0(t^-)}Pk_h(H)Y(t)}{Pk_h(H)} + o_p^{(h,t)}(1).$$

By Assumption 5, $Pk_h(H)Y(t) \geq MPk_h(H)$. Then $\frac{\frac{1}{S_0(t^-)}Pk_h(H)Y(t)}{Pk_h(H)} \geq \frac{\frac{1}{S_0(t^-)}MPk_h(H)}{Pk_h(H)} = \frac{M}{S_0(t^-)}$. Since $S_0(t^-)$ is bounded away from 0, we have $\frac{\frac{1}{S_0(t^-)}\mathbb{P}_n k_h(H)Y(t)}{\mathbb{P}_n k_h(H)}$ is bounded away from 0.

Let $A^h(u) = \frac{1}{S_0(u^-)} \frac{\mathbb{P}_n k_h(H) Y(u)}{\mathbb{P}_n k_h(H)}$ and $B^h(u) = \frac{\mathbb{P}_n k_h(H) N_j(u)}{\mathbb{P}_n k_h(H)}$. Let $B_0^h(t) = \frac{P k_h(H) N_j(t)}{P k_h(H)}$ and $A_0^h(t) = \frac{1}{S_0(t^-)} \frac{P k_h(H) Y(t)}{P k_h(H)}$. Then our point of interest is $(B, A) = (B_0^h(t), A_0^h(t))$. The components of the second term of our estimator $\widehat{F}_n(t \mid h)$ in (S14) can be mapped twice:

$$(B, A) \mapsto (B, \frac{1}{A}) \mapsto \int_{[0,t]} \frac{1}{A^h(u)} \, dB^h(u).$$

Using Lemma 12.3 and Lemma 6.19 from Kosorok (2008), this composition map is Hadamard differentiable on a domain of type $\{(B, A) : \sup_{\mathrm{h} \in \mathcal{H}} \int_{[0,\tau]} |dB^h(t)| \leq M,$ $\inf_{\substack{t \in [0,\tau] \\ h \in \mathcal{H}}} |A^h(t)| \geq \epsilon\}$ for all $h \in \mathcal{H}$ and for a given $M < \infty$ and $\epsilon > 0$, at every point $(B, A)$ such that $\frac{1}{A^h(t)}$ has bounded variation. Thus, the map is continuous. Using Assumption 9, the total variation of the kernels is bounded. Since we have shown that $A^h(t)$ and $B^h(t)$ are consistent, and $A^h(t)$ is bounded below by 0 and the map $\frac{1}{A^h(t)}$ is smooth, we apply the continuous mapping theorem so that the continuous map is also consistent.

Thus,

$$\sup_{\substack{t \in (0,\tau] \\ h \in \mathcal{H}}} \left| \widehat{F}_n(t \mid h) - F_j(t \mid h) \right| \to 0$$

in probability as $n \to \infty$.

□

# F Additional Simulation Details

The priority and competing event failure time hazard coefficients for both simulation settings are provided in Table S2.

Table S2: Cause-specific failure time hazard coefficients and covariate dimensions in Exponential and Fine-Gray simulation scenarios

| Model | | Scenario 1 (low dimension) | Scenario 2 (high dimension) |
|---|---|---|---|
| Exponential | $p$ | 3 | 6 |
| | $\beta_1^0$ | $(0, -1.5, -1.9, -0.3)^T$ | $(0, -0.5, -1.2, 1.5, 0.1, 0.6)^T$ |
| | $\beta_1^1$ | $(0, -0.6, -1.9, 1.5)^T$ | $(0, 0.3, -0.4, -0.2, 0.9, 1.6)^T$ |
| | $\beta_2^0$ | $(0, 1, -0.1, 0.4)^T$ | $(0, -0.6, 1.4, 1.5, 2, 1)^T$ |
| | $\beta_2^1$ | $(0, 1.2, 0.2, -1)^T$ | $(0, 1.5, -2.1, -0.8, 0.1, 1)^T$ |
| Fine-Gray | $p$ | 3 | 6 |
| | $\beta_1^0$ | $(0, 0.1, 0.3)^T$ | $(0, 0.9, -0.7, 0.6, -0.1, 0.5)^T$ |
| | $\beta_1^1$ | $(0, -1.8, -1.5)^T$ | $(0, -1.2, 0.1, -0.5, -0.2, -1.2)^T$ |
| | $\beta_2^0$ | $(0, -1.1, -0.3)^T$ | $(0, -0.5, -0.4, -1.1, -0.8, -0.4)^T$ |
| | $\beta_2^1$ | $(0, -0.2, 1.2)^T$ | $(0, 0.4, -0.3, 0.6, 0.2, 1.1)^T$ |

## Simple Exponential Simulation Setting

Since $T$ follows an exponential distribution, $S(t) = \exp\{-(\lambda_1^A + \lambda_2^A)t\}$ by definition of survival function. We derive $Pr(T \leq t_0, \epsilon = 1|Z_i)$ below using the property of independence as well as $1 = Pr(\epsilon = 1) + Pr(\epsilon = 2) = \frac{\lambda_1^A}{\lambda_1^A+\lambda_2^A} + \frac{\lambda_2^A}{\lambda_1^A+\lambda_2^A}$, and the derivation for $Pr(T \leq t_0, \epsilon = 2|Z_i)$ is defined similarly.

$$
\begin{aligned}
Pr(T \le t_0, \epsilon = 1|Z_i) &= Pr(T_1 \le t_0, T_2 \le t_0, T_1 < T_2) \\
&= Pr(T_1 < T_2 \le t_0) \\
&= \int_0^{t_0} \int_0^{t_2} f(t_1, t_2) dt_1 dt_2 \\
&= \int_0^{t_0} \int_0^{t_2} f(t_1) f(t_2) dt_1 dt_2 \\
&= \int_0^{t_0} \int_0^{t_2} \lambda_1^A \exp(-\lambda_1^A t_1) \lambda_2^A \exp(-\lambda_2^A t_2) dt_1 dt_2 \\
&= \int_0^{t_0} \lambda_2^A \exp(-\lambda_2^A t_2) \int_0^{t_2} \lambda_1^A \exp(-\lambda_1^A t_1) dt_1 dt_2 \\
&= \int_0^{t_0} \lambda_2^A \exp(-\lambda_2^A t_2)[1 - \exp(-\lambda_1^A t_2)] dt_2 \\
&= \lambda_2^A \left( \int_0^{t_0} e^{-\lambda_2^A t_2} dt_2 - \int_0^{t_0} e^{-(\lambda_1^A + \lambda_2^A) t_2} dt_2 \right) \\
&= -\lambda_2^A \left( -\frac{1}{\lambda_2^A} (1 - e^{-\lambda_2^A t_0}) \right) - \lambda_2^A \left( \frac{1}{\lambda_1^A + \lambda_2^A} (1 - e^{-(\lambda_1^A + \lambda_2^A) t_0}) \right) \\
&= \left( 1 - e^{-\lambda_2^A t_0} \right) - \frac{\lambda_2^A}{\lambda_1^A + \lambda_2^A} \left( 1 - e^{-(\lambda_1^A + \lambda_2^A) t_0} \right) \\
&= 1 - e^{-\lambda_2^A t_0} - \frac{\lambda_2^A}{\lambda_1^A + \lambda_2^A} + \frac{\lambda_2^A}{\lambda_1^A + \lambda_2^A} \left( e^{-(\lambda_1^A + \lambda_2^A) t_0} \right) \\
&= 1 - \frac{\lambda_2^A}{\lambda_1^A + \lambda_2^A} - e^{-\lambda_2^A t_0} + \frac{\lambda_2^A}{\lambda_1^A + \lambda_2^A} \left( e^{-(\lambda_1^A + \lambda_2^A) t_0} \right) \\
&= \frac{\lambda_1^A}{\lambda_1^A + \lambda_2^A} - e^{-\lambda_2^A t_0} + \frac{\lambda_2^A}{\lambda_1^A + \lambda_2^A} \left( e^{-(\lambda_1^A + \lambda_2^A) t_0} \right).
\end{aligned}
$$

## Fine-Gray Simulation Setting

When covariates are 0, $\Pr(T_i \le t, \epsilon = 1|Z_i) = p\{1 - \exp(-t)\}$, where $1 - \exp(-t)$ is the cumulative distribution function (CDF) of exp(1). Because of how we define the mixture above, $\Pr(T_i < \infty, \epsilon = 1) = p$. Thus, $\Pr(T_i = \infty, \epsilon = 1) = 1 - p$. Then, $\Pr(\epsilon = 1|Z_i) = \Pr(T < \infty, \epsilon = 1|Z_i) = \lim_{t \to \infty} 1 - [1 - p(1 - \exp(-t))]^{\exp(Z_i^T \beta_1(A))} = 1 - [1 - p]^{\exp(Z_i^T \beta_1^A)}$.

To simulate cause 1 failure time, we use the property that the distribution of a CDF of a random variable is uniform. We generate an uniform random variate $u_1$ from Uniform(0,1) distribution and set this equal to the CDF of failure time given cause 1 and covariates. By definition of conditional probability, $\Pr(T_i \le t|\epsilon = 1, Z_i) = \frac{\Pr(T_i \le t, \epsilon = 1|Z_i)}{\Pr(\epsilon = 1|Z_i)}$, so

$$
\Pr(T_i \le t|\epsilon = 1, Z_i) = \frac{1 - [1 - p\{1 - \exp(-t)\}]^{\exp(Z_i^T \beta_1^A)}}{1 - (1 - p)^{\exp(Z_i^T \beta_1^A)}} \overset{\text{set}}{=} u_1.
$$

We derive failure time from cause 1, $t_1$, by solving for $t$.

$$\begin{aligned}
\frac{1-[1-p(1-\exp(-t))]^{\exp(Z_i^T\beta_1^A)}}{1-(1-p)^{\exp(Z_i^T\beta_1^A)}} &= u_1 \\
1-[1-p(1-\exp(-t))]^{\exp(Z_i^T\beta_1^A)} &= u_1 \times \left(1-(1-p)^{\exp(Z_i^T\beta_1^A)}\right) \\
[1-p(1-\exp(-t))]^{\exp(Z_i^T\beta_1^A)} &= 1-\left(u_1 \times \left(1-(1-p)^{\exp(Z_i^T\beta_1^A)}\right)\right) \\
\exp(Z_i^T\beta_1^A) \times \log\left(1-p(1-\exp(-t))\right) &= \log\left(1-u_1 \times \left(1-(1-p)^{\exp(Z_i^T\beta_1^A)}\right)\right) \\
\log\left(1-p(1-\exp(-t))\right) &= \frac{\log\left(1-u_1 \times \left(1-(1-p)^{\exp(Z_i^T\beta_1^A)}\right)\right)}{\exp(Z_i^T\beta_1^A)} \\
1-p(1-\exp(-t)) &= \exp\left(\frac{\log\left(1-u_1 \times \left(1-(1-p)^{\exp(Z_i^T\beta_1^A)}\right)\right)}{\exp(Z_i^T\beta_1^A)}\right).
\end{aligned}$$

This implies

$$\begin{aligned}
\exp(-t) &= 1-p^{-1} \times \left(1-\exp\left(\frac{\log\left(1-u_1 \times \left(1-(1-p)^{\exp(Z_i^T\beta_1^A)}\right)\right)}{\exp(Z_i^T\beta_1^A)}\right)\right) \\
t &= -\log\left(1-p^{-1} \times \left(1-\exp\left(\frac{\log\left(1-u_1 \times \left(1-(1-p)^{\exp(Z_i^T\beta_1^A)}\right)\right)}{\exp(Z_i^T\beta_1^A)}\right)\right)\right) \\
\implies t_1 &= t.
\end{aligned}$$

The CIF for failures from cause 2 was obtained by taking $\Pr(\epsilon = 2|Z_i) = 1 - \Pr(\epsilon = 1|Z_i) = 1-(1-[1-p]^{\exp(Z_i^T\beta_1^A)}) = (1-p)^{\exp(Z_i^T\beta_1^A)}$. We use the exponential distribution with rate $\exp(Z_i^T\beta_2^A)$ for $\Pr(T_i \leq t|\epsilon = 2, Z_i)$. In other words, $t_2 \sim \exp(\exp(Z_i^T\beta_2^A))$. Without loss of generality, we could also derive $t_2$ similarly to $t_1$ (see Supplementary Material). Censoring time $C_i$ follows a Uniform$(0, \tau = 4)$ distribution, independent of failure time.

Using the assumptions or derivations established earlier, the overall survival function when the priority cause is cause 1 can be written as:

$$\begin{aligned}
S(t) = \Pr(T > t) &= 1 - \Pr(T \leq t) = 1 - \left\{ \Pr(T \leq t, \epsilon = 1) + \Pr(T \leq t, \epsilon = 2)\right\} \\
&= 1 - \left\{ \Pr(T \leq t, \epsilon = 1) + \Pr(T \leq t|\epsilon = 2) \times \Pr(\epsilon = 2)\right\} \\
&= 1 - \Big\{1 - [1-p\{1-\exp(-t)\}]^{\exp(Z_i^T\beta_1^A)} \\
&\qquad + (1-\exp(-\exp(Z_i^T\beta_2^A) * t)) \times (1-p)^{\exp(Z_i^T\beta_1^A)}\Big\} \\
&= [1-p\{1-\exp(-t)\}]^{\exp(Z_i^T\beta_1^A)} - (1-\exp(-t \times \exp(Z_i^T\beta_2^A))) \times (1-p)^{\exp(Z_i^T\beta_1^A)}.
\end{aligned}$$

We derive $t_2$ from the Fine-Gray simulation setting in a similar manner as $t_1$. Since $\Pr(\epsilon = 2|Z_i) = 1-\Pr(\epsilon = 1|Z_i)$, $\Pr(\epsilon = 2|Z_i) = 1-(1-[1-p]^{\exp(Z_i^T\beta_1^A)}) = [1-p]^{\exp(Z_i^T\beta_1^A)}$. Then, we have

$$\begin{aligned}
\Pr(T_i \leq t, \epsilon = 2|Z_i) &= \Pr(T_i \leq t|\epsilon = 2, Z_i) \times \Pr(\epsilon = 2|Z_i) \\
&= (1-\exp(-\exp(Z_i^T\beta_2(A))t)) \times (1-p)^{\exp(Z_i^T\beta_1^A)}.
\end{aligned}$$

Setting this to a random variate $u_2 \sim Uniform(0,1)$, solve for $t_2$ to obtain failure time for cause 2.

$$\begin{aligned}
\text{Set } u_2 &= (1-\exp(-\exp(Z_i^T\beta_2^A)t_2)) \times (1-p)^{\exp(Z_i^T\beta_1^A)} \\
\frac{u_2}{(1-p)^{\exp(Z_i^T\beta_1^A)}} &= 1-\exp(-\exp(Z_i^T\beta_2^A)t_2) \\
1-\frac{u_2}{(1-p)^{\exp(Z_i^T\beta_1^A)}} &= \exp(-\exp(Z_i^T\beta_2^A)t_2) \\
\log\left(1-\frac{u_2}{(1-p)^{\exp(Z_i^T\beta_1^A)}}\right) &= -\exp(Z_i^T\beta_2^A)t_2 \\
t_2 &= -\frac{\log\left(1-\frac{u_2}{(1-p)^{\exp(Z_i^T\beta_1^A)}}\right)}{\exp(Z_i^T\beta_2^A)}.
\end{aligned}$$

# G Peripheral Artery Disease Cohort Details

We estimated the criterion $\phi$ for each policy $d$ as:

$$\hat{\phi}_{\text{OS}}(d) = \frac{\sum_{i:\text{test set}}(T_i \wedge \tau)W_i^{\text{OS}}(d)}{\sum_{i:\text{test set}} W_i^{\text{OS}}(d)} \text{ and } \hat{\phi}_{\text{PC}}(d) = \frac{\sum_{i:\text{test set}}(T_{1i} \wedge \tau)W_i^{\text{PC}}(d)}{\sum_{i:\text{test set}} W_i^{\text{PC}}(d)}, \qquad \text{(S22)}$$

where $W_i^{\text{OS}}(d) = \frac{I(d(z_i)=A_i)\delta_{0i}}{\hat{Pr}(A_i|z_i)\hat{Pr}(C_i>T_i|z_i)}$ and $W_i^{\text{PC}}(d) = \frac{I(d(z_i)=A_i)\delta_{1i}}{\hat{Pr}(A_i|z_i)\hat{Pr}(C_i \wedge T_{2i}>T_{1i}|z_i)}$. Recall that $\delta_{1i} = I(T_{1i} \leq C_i \wedge T_{2i})$. Following Cho et al. (2023), we used the test set to estimate the propensity scores and censoring probabilities via random forests and random survival forests, respectively, adjusting for the same covariates as in the ITR estimation. However, in the propensity score model, treatment is the outcome rather than a predictor. We performed 10-fold cross-validation (CV) to avoid overfitting ($K_{\text{cv}} = 10$). We split each CV fold into training (80%) and testing (20%) datasets. We set $\alpha_\phi = 0.1$, and set the number of trees to 300, minimum node size to 5, and minimum number of events to be 2.